\documentclass[]{scrartcl}

\usepackage{amsmath}
\usepackage{amssymb}
\usepackage{slashed}
\usepackage{color}
\usepackage{makeidx}
\usepackage{tikz}
\usepackage{cite}
\usepackage[normalem]{ulem}
\usetikzlibrary{decorations.pathmorphing}
\usetikzlibrary{decorations.markings}
\usetikzlibrary{calc}
\title{A primer on Higgs Effective Field Theory with Geometry}
\author{Rodrigo Alonso\\
	\textit{\large Institute for Particle Physics Phenomenology, Durham University}
}
\makeindex

\begin{document}

\maketitle

\begin{abstract}
These lecture notes, prepared for the 2022 QUC summer school at KIAS, provide an introduction to Higgs Effective Field Theory and the use of field geometry in Quantum Field Theory. While not sounding the depths of any of these topics, we will cover and give a sense of the inner workings of: the action for Goldstone bosons, the independence of scattering amplitudes from field parametrisations, linear vs non-linear realizations --their `geography'  and experimental prospects to tell them apart--, ultra-violet completions and the LSZ formula for fields in curved space.
\qquad\qquad\qquad\quad
{\small Preprint: IPPP/23/38}
\end{abstract}

\tableofcontents
\section{Introduction}

The foundation for our description of elementary particle physics is Quantum Field Theory; within this framework, 
one is offered two paths to describe Nature: Models and Effective field theory (EFT). In broad strokes and contrast with one another one can say of each that  
\begin{itemize}
	\item Models are --when compared to the alternative-- predictive, complete and have some \textit{raison d'\^etre}; i.e. a motivation or question they try to address.
	\item EFTs are --when compared to the alternative-- general, applicable in a limited regime and unbiased. 
\end{itemize}

Typically Models are formulated in answer to open questions in physics and offer some deeper knowledge about Nature; examples in particle physics include supersymmetry, extra-dimensional and composite Higgs models to address the hierarchy problem, Froggatt Nielsen or discrete symmetries to shed light on the flavour puzzle, axions for the strong CP problem, countless dark matter candidates to account for the missing matter component of the universe etc etc. However, when faced with a vast literature on models trying to answer a given question, one might wonder which one to choose given all of them but one --or perhaps all of them-- are wrong. The alternative of EFT does not offer answers to deep questions but it does help 
with decision fatigue: it uses the minimal set of assumptions while encompassing many possibilities at once. All EFTs require for their formulation is 
\begin{enumerate}
	\item\label{ITMA} The particle (field) content
	\item\label{ITMB} The symmetries
	\item\label{ITMC} An expansion parameter
\end{enumerate}
In fact when spelled like that, one might argue that these are the minimum requirements for doing physics in any situation.

It is this short list of demands of EFTs that implies the generality in their description of Nature; any dynamics compatible with the items in this list is included in the EFT. This generality on the other hand might come with a large theory space, cumbersome to deal with as a whole, but typically much reduced when one is interested in a given experimental process. The finite regime of validity mentioned in the comparison with models is related to the expansion parameter that organizes theory predictions; this parameter is a function of the variables characterising a physical process --typically a ratio of scales-- and when it ceases to be small we have found the `breaking point' of the EFT. It is at this point that some other more complete description of Nature takes over.
One last structural remark is that, within its validity regime, there is no obstacle to precise quantum-level computations in an EFT and, in the jargon of QFT, non-renormalisable theories are renormalisable to any finite order in the expansion. 

In these lecture notes the physics under inspection is the electroweak interaction which is currently being probed experimentally at the Large Hadron Collider (LHC). The exploration of this frontier --and much of it remains in the dark-- will be one of the main legacies of the LHC. To learn from and interpret the experimental data the theory framework should be sufficiently developed and general to cover the most possibilities. EFT, in view of the properties we outlined above, is well suited for the job; the development of the most general EFT for electroweak physics lead to what is now known as Higgs Effective Field Theory (HEFT). Provided the mass of new particles is well separated from the electroweak scale,~$v=246$GeV, HEFT is the most general possible description of physics. The exploration and construction of this encompassing EFT has brought new theory insights into UV completion, amplitudes, and geometry; these notes aim at giving a starting point to understand this progress and push it forward.

The organisation of these notes is as follows: sec.~\ref{sec:Long} sets out the procedure to describe a Goldstone action and applies it to the electro-weak theory. Sec.~\ref{sec:Hggs}  introduces the Higgs particle and HEFT together with the limit that produces the Standard Model. Sec.~\ref{sec:Exp} discusses the expansion parameter, while sec.~\ref{sec:Geo} introduces geometry for field space, sec.~\ref{sec:Dic} presents the linear non-linear dichotomy and sec.~\ref{sec:PH} discusses the phenomenology.

Finally let me say that the list of references, as ever, is incomplete and in particular in the year that elapsed between the writing of this notes and their placement on the arXiv there has been notable progress in the field, a sample being~\cite{Helset:2022pde,Helset:2022tlf,Gomez-Ambrosio:2022why,Sun:2022ssa,Cohen:2022uuw,Graf:2022rco}.

\section{Longitudinal degrees of freedom}\label{sec:Long}

 One of the uses of EFT is a bottom-up approach to unknown physics. In our present case of study the bottom is low energy and the up is high energy, but in other instances this bottom-up arrow might be pointing in other directions.
 In this spirit we will follow the \{\ref{ITMA}\},\{\ref{ITMB}\},\{\ref{ITMC}\} steps enumerated in the introduction to specify an EFT; so we start with the particle content and symmetries. 
 
 The gauge symmetry at the heart of the known interactions, gravity excluded, is
 \begin{align}
 \mathcal G=SU(3)_c\times SU(2)_L\times U(1)_Y\label{Gaug}\,,
 \end{align}
 and hence comes with dim(Lie Algebra)$=8+3+1$ spin one massless particles; let us call them $G_\mu$, $W_\mu$ and $B_\mu$. In addition the matter content is specified by the representation under the group in eq.~(\ref{Gaug}),
 
 \begin{align} \nonumber
 	&&&q_L &&q_R&& \ell_L&& \ell_R\\\hline \nonumber
 	&SU(3)_c&&\mathbf{3}&& \mathbf{3}&& -&& -\\ \nonumber
 	&SU(2)_L&&\mathbf{2}&& -&& \mathbf{2}&& -\\
 	&U(1)_Y&&\frac{1}{6}&& \frac16+\frac{\sigma_3}{2}&& -\frac12&& -\frac12+\frac{\sigma_3}{2}\label{U1Y}
 \end{align}
 where $\mathbf{N}$ is the fundamental representation of ($SU(N)$) for $\mathbf{3},\mathbf{2}$, $R$ \& $L$ stand for right and left handed fermions respectively, $\sigma_3$ is the third Pauli matrix ($\sigma_3=$Diag$(1,-1)$) and we find it useful to group right-handed fields in doublets $q_R=(u_R,d_R)$, $\ell_R=(0,e_R)$ -- feel free to add a RH neutrino if you would prefer that.
 
 This table gives the transformation properties of each field, so if we describe such transformation with an element of the group $G$ ($G^\dagger G=1$) the transformed quark doublet $(q_L)_G$ is 
 \begin{align}\label{GqL}
  (q_L)_G&\equiv Gq_L=(1+i\epsilon\cdot T )q_L+\mathcal O(\epsilon^2) \,,
 	&&\delta_G q_L\equiv i\epsilon \cdot T q_L\,,\\
  (\partial_\mu q_L)_G&=(\partial_\mu G) q_L+G\partial_\mu q_L \,,
 	&&\label{GagmQ}
 \end{align}
 where $T$ are the generators of the gauge symmetry in eq.~(\ref{Gaug}) and $\epsilon$ parametrizes the transformation ($G=e^{i\epsilon \cdot T}$) and has an index that runs over the generators $\epsilon\cdot T=\epsilon^\alpha T_\alpha$ -- we use the repeated index summation convention but note these are not Lorentz indexes. One has that the field transforms covariantly but its derivative does not, to solve it one introduces 
 $D_\mu=\partial_\mu+igA_\mu$to maintain
 \begin{align}\label{GaugG}
 	(D_\mu q)_G&=GD_\mu q\,, &&\textrm{with}  &(ig A_\mu)_G=&G\partial_\mu G^\dagger +GigA_\mu G^\dagger\,.
 \end{align}
   Explicitly for our quark doublet
 \begin{align}
 D_\mu q_L = \left(\partial_\mu+ig_s\frac{\lambda_a}{2}G_\mu^a+ig\frac{\sigma_I}{2}W_\mu^I+\frac{ig'}{6}B_\mu\right)q_L\,,
 \end{align}
 where $\lambda_a$ are the Gellmann matrices, $\sigma^I$ the Pauli matrices and $G$, $W$ and $B$ are the gauge bosons of $SU(3)_c$, $SU(2)_L$ and $U(1)_Y$ respectively with couplings $g_s$, $g$ and $g'$.
 
 We have lastly to mention that not all this symmetry is explicit at low energy, of the electroweak group only $U(1)_{\textrm{em}}$ survives. This breaking allows for the masses of fermions and requires masses for gauge bosons. A massive vector boson requires a longitudinal component not present in the gauge theory, so one introduces Goldstone bosons which therefore `live' in 
$$
\textrm{Goldstone space}\qquad\quad SU(2)_L\times U(1)_Y /U(1)_{\textrm{em}} \,.
$$
How does one go about describing fields living in such a space?

\subsection{Goldstone Bosons}
 The procedure we use to describe Goldstone bosons here is known as CCWZ after the authors of the seminal papers~\cite{Coleman:1969sm,Callan:1969sn}. The essential tool for the formulation of this procedure is group theory, as applied to the breaking of a group $\mathcal G$ down to an unbroken subgroup $\mathcal H$, denoted schematically as $\mathcal G\to \mathcal H$, and with the rest of the group $\mathcal G/\mathcal H$ referred to as coset. The derivation as here presented applies to any compact group and is general; it will in turn require some abstraction, more than is needed for HEFT, so any student eager to get there can skip this section and move to~\ref{EWGB}. 
 
 Let us discuss the Lie algebra by denoting generators as follows
 \begin{center}
 \begin{tabular}{c|c|c|c}
 	space & $\mathcal G$ & $\mathcal H$ & $\mathcal G/\mathcal H$\\\hline
 	Generators & $T_\alpha$ & $t_a$ & $X_A$
 \end{tabular}
\end{center}
so  $\{T\}=\{X,t\}$, and we use Greek indexes for the encompassing  $T$-generators, Latin upper case and lower case for the broken and unbroken respectively. The generator normalization chosen and Lie algebra are (here $T^\dagger=T$ and $f_{\alpha\beta}{}^{\gamma}$ are the structure constants, $[T_{\alpha},T_{\beta}]=if_{\alpha\beta}{}^{\gamma}T_\gamma$) 
\begin{align}\label{Lie}
[t_a,t_b]=i &f_{ab}{}^{c}t_c\,, & \textrm{Tr}\left(t_a X_A\right)&=0\,,\\
[t_a,X_B]=i&f_{aB}{}^{C} X_C\,, & \textrm{Tr}\left(X_A X_B\right)&=\delta_{AB}\,,\\
[X_A,X_B]=i&f_{AB}{}^{c} t_c+if_{AB}{}^{C} X_C\,, & \textrm{Tr}\left(t_a t_b\right)&=\delta_{ab}\,.
\end{align}
The absence of $X$ on the RHS of the first commutator equation follows from $\mathcal H$ being closed --otherwise it wouldn't define a group-- whereas the absence of $t$ on the second follows from the above statement and the fully anti-symmetric structure constants of a compact group. The form of these two commutators will be useful later.

 Next we point out that any element of the group $G_T$  can be factorized into an element of the broken part, or coset, and the unbroken group as
\begin{align}\label{Fact}
	G_T=&G_X G_t\,, & G_t&\in \mathcal H\,, & G_X&\in \mathcal G/\mathcal H\,,
\end{align}
and one can parametrize an element of the coset, $\xi(x)$, with an exponential representation; we write it in terms of dim$\mathcal(G/\mathcal H)$ fields $\pi^A(x)$ as
\begin{align}
	\xi&\in \mathcal G/\mathcal H\,,  & \xi=&e^{i\bar\pi}=e^{i\pi\cdot X/v}\,,
\end{align}
where $v$ is the decay constant needed for the right mass dimension of our scalar fields $\pi^A$. The transformation of $\xi$ follows from the composition rule of the group; if we apply two transformations $G_1$, $G_2$ subsequently, we obtain a transformation $G_2G_1$ hence, regarding $\xi$ as our first transformation and $G$ as the second,
\begin{align}
	\xi &\to G \xi\equiv\xi_G G_t \,,& \xi_G&= G\xi G_t^\dagger(\xi)\equiv\xi+i\epsilon\cdot T\,\xi-i\xi\, \epsilon\cdot\beta(\xi)\cdot t+\mathcal O(\epsilon^2)\,,\label{xiG}
\end{align}
where $\epsilon$ parametrises the transformation as in eq.~(\ref{GqL}), we have defined the function of $\xi$, $\beta(\xi)$, which has two indexes $\beta_{\alpha a}$, and to find the transformed $\xi$, $\xi_G$, we have used again the factorization of eq.~(\ref{Fact}).

The Goldstone-dependent unbroken group element $G_t(\xi)$ is defined implicitly so that $G\xi G^\dagger_t$ is an coset element. Given it is not something you might be used to, it is good to be more explicit and write, with $(\bar\pi)_G=\bar\pi+\delta_G\bar\pi+\mathcal O(\epsilon^2)$
\begin{align}\label{dxiG2}
	\xi^\dagger\delta_G \xi=e^{-i\bar\pi}\delta_G \pi^A\frac{\partial}{\partial\pi^A} e^{i\bar\pi}=&e^{-i\bar\pi} i\epsilon\cdot T e^{i\bar\pi}-i\epsilon\cdot \beta(\bar\pi)\cdot t\,,\\ \label{dxiG}
\sum	\frac1{n!}\underbrace{[ -i\bar\pi\, , [-i\bar\pi\,\dots\, [-i\bar\pi}_{n-1 \textrm{~times}} , i\delta_G\bar\pi]...]]=&\sum\frac{1}{n!}\underbrace{[ -i\bar\pi\, , [-i\bar\pi\,\dots\, [-i\bar\pi}_{n \textrm{~times}}, i\epsilon\cdot T ]...]]-i\epsilon\cdot\beta \cdot t \,,
\end{align}
where terms with an underbrace in the last line mean nested commutators\footnote{The expression in~(\ref{dxiG}) follows from~(\ref{dxiG2}) via use of  Baker-Campbell-Hausdorff formula.} and we have taken $\xi^\dagger$ times the first variation,  $\xi^\dagger\delta_G\xi$, since this construction belongs in the Lie algebra -- if this is not evident you can use the second line above to derive this result.
The above equation can be cast, to the second order level in $\bar \pi$, as

 {\noindent\color{teal}{\rule{\textwidth}{1mm}}}
  \textbf{Do it yourself \#1:} Finding the transformation rule for $\pi$ to   $\mathcal O(\pi^2)$. Expand eq.~(\ref{dxiG}) to $\mathcal O(\pi^2)$ on both sides including $\beta(\bar\pi)$, then multiply on the left by a matrix $1+a_L\bar\pi$  and on the right by $1+a_R\bar\pi$  with $a_{L,R}$ such that only the first term, $i\delta_G\bar\pi$, is left on the LHS; then simplify the RHS to find eq.~(\ref{XiGexp}). Finally solve for $\beta(0)$, $\partial\beta(0)$  such that the RHS projects only on the broken Lie algebra and use the commutation relations of~(\ref{Lie}) to find eq.~(\ref{XiTExp}).
   
  {\noindent\color{teal}{\rule{\textwidth}{1mm}}}
\begin{align}\label{XiGexp}
	i\delta_G\bar\pi=i\epsilon \cdot T-i\epsilon\cdot\beta(0) \cdot t+\frac{1}{2}\left[\bar\pi,\epsilon\cdot T\right]+\frac{1}{2}\left[\bar\pi,\epsilon\cdot\beta(0)\cdot t\right]-i\epsilon\cdot\left(\pi^A\frac{\partial\beta}{\partial\pi^A}\right)\cdot t\,,
\end{align}
The LHS belongs in the Lie algebra of the coset --i.e. there's only $X$'s; it is the defining job of $\beta$ to ensure the RHS does too, so solving for the equation above gives the explicit form of $\beta$. Once this form is found and put back into eq.~(\ref{XiGexp}) we obtain the transformation rule for our Goldstones to first order in $\pi$ as 
\begin{align}\label{XiTExp}
	\delta_G\pi^A=v\epsilon^{A}_X+f_{B a}{}^{A}\pi^B\epsilon^a_t+\frac12 f_{BC}{}^{A}\pi^B\epsilon^C_X+\mathcal{O}(\pi^2)\,,
\end{align}
where we have split $\epsilon$ in broken ($\epsilon_X$) and unbroken ($\epsilon_t$) group components $\epsilon\cdot T=\epsilon_X \cdot X+\epsilon_t\cdot t$. One can see that a transformation generated in the broken Lie algebra $X$ shifts the fields by $v\epsilon_X$ as is the case for axions, but if the group is non-abelian there will also be higher orders in $\pi$ given by the structure constants. On the other hand a unbroken group transformation $\epsilon_t$ does not shift the fields but starts at the linear level.

This gives us a sense of the group action on the Goldstones but it is expanded on $\bar\pi$ and we want our action invariant under the full non-linear transformation. No invariant operator can be built with $\xi$ alone --e.g. $\xi^\dagger \xi=1$-- and hence no potential is present. One turns to derivatives next to encounter that the transformation here is $x$-dependent on two fronts, $\epsilon$ and $G_t(\xi)$, as opposed to $q_L$, so
\begin{align}
	(\xi^\dagger \partial_\mu \xi)_G=&G_t \xi^\dagger (\partial_\mu \xi) G_t^\dagger+G_t \xi^\dagger G^\dagger (\partial_\mu G) \xi G_t^\dagger+G_t^\dagger(\xi) \partial_\mu G_t^\dagger(\xi)\\\nonumber
	=&G_t \xi^\dagger (\partial_\mu \xi) G_t^\dagger+G_t \xi^\dagger \left(G^\dagger\partial_\mu \epsilon\frac{\partial}{\partial\epsilon} G\right) \xi G_t^\dagger+G_t(\xi)\left( \partial_\mu\epsilon\frac{\partial}{\partial\epsilon}+\partial_\mu\pi^A\frac{\partial}{\partial\pi^A}\right) G_t^\dagger(\xi)\,.
\end{align}
The last two terms keep this construction from having covariant transformation properties. Dealing with the second term, $G^\dagger \partial_\mu G$, is not anything new and we can do as in eq.~(\ref{GagmQ}), yet the last one, $G_t(\xi)\partial_\mu G_t^\dagger(\xi)$, is of a different kind. This new term, we first note, affects only the projection on the unbroken Lie algebra since
\begin{align}
\textrm{Tr}\left(X_A G_t(\xi)\partial_\mu G^\dagger_t(\xi)\right)=0  \,,
\end{align}
as follows from~(\ref{Lie}). It seems convenient then to separate $\xi^\dagger D_\mu\xi$ in the two sectors of our Lie algebra and reassess: 
\begin{align}\label{JVDef}
J_A&\equiv\textrm{Tr}\left(X_A \xi^\dagger D_\mu \xi\right)=\textrm{Tr}\left(X_a \xi^\dagger \left(\partial_\mu+igA\cdot T\right) \xi\right)\,,&
V_a&\equiv\textrm{Tr}\left(t_a \xi^\dagger D_\mu \xi\right)\,,
\end{align}
in terms of which the transformation properties are 
\begin{align}
	\left(J_A\right)_G= \textrm{Tr}\left(X_A G_t\left(J^BX_B+V^at_a\right)G_t^\dagger \right)=\textrm{Tr}\left(X_A G_t\left(J^BX_B\right)G_t^\dagger \right)\,,
\end{align}
since, again using the exponential commutator formula and eq.~(\ref{Lie}), one can see that $G_tXG_t^\dagger$ maps back to $X$ and $G_ttG_t^\dagger$ to $t$. The analogous treatment for $V$ shows that 
\begin{align}
	(J\cdot X)_G&=G_t (J\cdot X)G^\dagger_t\,, &
	(V\cdot t)_G&=G_t (V\cdot t)G^\dagger_t+G_t(\xi)\partial_\mu G^\dagger_t(\xi)\,.
\end{align}
 We have encountered in fact a different kind of gauge transformation via the gauge-boson structure $V$ that transforms with a shift. Building covariant transforming combinations out of $V$ then mimics a pure gauge theory; one can define a field strength and its derivatives, see e.g.~\cite{Contino:2015mha}.

We are interested nonetheless on the lowest derivative term in the Lagrangian, since this will be leading in the EFT expansion --a point to be expanded on in sec.~\ref{sec:Exp}. This operator is the kinetic term for the Goldstones, and reads
\begin{align}\label{GBLag}
\mathcal L_{p^2}= -\frac{v^2}{2} \textrm{Tr}\left(J_\mu\cdot X J^\mu\cdot X\right)=-\frac{v^2}{2}\textrm{Tr}\left(X_a \xi^\dagger D_\mu \xi\right)\textrm{Tr}\left(X_a \xi^\dagger D_\mu \xi\right)\,.
\end{align}

Let us apply next this procedure to the electro-weak theory; we will find that a short-cut allows reaching the Lagrangian avoding some of the derivation above.
 
 \subsection{Electro-weak Goldstone Bosons}\label{EWGB}
It is generally a useful approach to exploit all symmetries of our system, even if they are only approximate. The scalar sector of the electroweak theory has a global symmetry that comprises the gauge, $SO(4)\sim SU(2)_L\times SU(2)_R$ and the breaking pattern can be embedded as 
$$
SU(2)_L \times SU(2)_R /SU(2)_V \,,
$$
where $SU(2)_L$ is the same as in eq.~(\ref{Gaug}), $SU(2)_R$ acts on $q_R$, $\ell_R$ and in the scalar sector contains $U(1)_{Y}$ whereas $SU(2)_V$ contains $U(1)_{\textrm{em}}$ --for the fermion sector $U(1)_Y$ is a combination of $SU(2)_R$ and $B-L$ as can  be seen in eq.~(\ref{U1Y}).
The generators for the coset are $4\times 4$ matrices as
\begin{align}\label{GensEW}
	X^I=\frac12\left(\begin{array}{cc}
	\sigma^I&\\&-\sigma^I
	\end{array}\right)\,,
\end{align}
with $\sigma$ the Pauli matrices. When exponentiated the form of $\xi$ will still be block diagonal, explicitly 
\begin{align}
\xi\equiv&\left(\begin{array}{cc}
U_L&\\&U_R
\end{array}\right), &
U_L^\dagger U_L=U_R^\dagger U_R\,,&=1\,,\qquad \det(U_L)=\det(U_R)=1\,,
\end{align}
and our group transformation reads, c.f.~(\ref{xiG}),
\begin{align} 
(\xi)_G=&G\xi G_t=\left(\begin{array}{cc}\mathbf L&\\& \mathbf R
\end{array}\right)\left(\begin{array}{cc}
U_L&\\&U_R
\end{array}\right)\left(\begin{array}{cc}
\mathbf V^\dagger(\xi)&\\
&\mathbf V^\dagger (\xi)
\end{array}\right)\,.
\end{align}
The complication in the description arises from the $\xi$-dependent transformation $\mathbf V$ on the right, not sure if you can spot it from here, but
the shortcut is to get rid of $\mathbf V$ by the following construction
\begin{align}
	  U(x)&\equiv U_L(x) U_R^\dagger (x)\,, &
	 (U)_{G}&=\mathbf LU\mathbf R^\dagger\,,
\end{align}
where crucially, given $\mathbf L$ and $\mathbf R$ do not depend on the fields $\xi$, $U$ transforms as an ordinary (linear) representation. This attains the same goal as the current $J$ since one can obtain a covariant derivative acting on $U$ in terms of gauge bosons $A_\mu$ as
\begin{align}
D_\mu U= \left(\partial_\mu+ig\frac{\sigma_I}2 W_\mu^I\right)U -ig'U \frac{\sigma_3}2B_\mu\,,
\end{align}
which transforms as $(D_\mu U)_G= \mathbf L (D_\mu U) e^{-i\sigma_3 \epsilon_Y/2}$.
Well then, let's put this together with the kinetic terms for the fermions and gauge bosons and a Goldstone-fermion coupling to find
\begin{align}\nonumber
	\mathcal L =& \frac{v^2}{4} \textrm{Tr}\left( D_\mu U D^\mu U^\dagger\right)+i\sum_\psi\left(\bar\psi_L\slashed D\psi_L+i\bar\psi_R\slashed D\psi_R\right)-\frac14 \sum_G F_{\mu\nu}F^{\mu\nu}\\&-\sum_\psi \frac{v}{\sqrt2}\bar\psi_L U \mathcal Y_\psi\psi_R+h.c.\,, \label{actU} 
\end{align}
where $\mathcal Y_q=$Diag$(Y_u,Y_d)$, $\mathcal Y_\ell=$Diag$(0,Y_e)$ each $Y_i$ a $3\times 3$ matrix in flavor space.
While the invariance of the first line terms might be easier to spot, it is a good exercise to check the invariance of the Yukawa terms on the second line for both quarks and leptons --in particular for hypercharge we have $e^{iQ_Y\epsilon_Y}$ with charge $Q_Y$ given in eq.~(\ref{U1Y}).

The Lagrangian in eq.~(\ref{actU}) gives a good description of physics below 100 GeV; when expanded around the vacuum, conventionally $\langle U\rangle =1$, we obtain masses as
\begin{align}
	m_\psi=&\frac{vY_i}{\sqrt2}\,, &
	M_W&=gv/2\,, & M_Z&=\frac{gv}{2c_{\theta_W}}\,,
\end{align}
with $\tan(\theta_W)=g'/g$, $Z=c_{\theta_W}W^3-s_{\theta_W}B$ and the photon field being the orthogonal combination. This theory in fact is as much as we knew for certain about nature before 2012, as I remember learning when sitting in a physics school in 2011 as you are now sitting on one too.

What if we do not assume custodial symmetry in the scalar sector? One need not preserve $SU(2)_R$ but only $U(1)_Y$ whose generator in scalar space corresponds to the third Pauli matrix in $SU(2)_R$. This means, since now we have an abelian group, that insertions of the hypercharge generator are allowed; this has historically been done with $\mathbf{T}=U\sigma_3 U^\dagger$ ($(\mathbf{T})_G=\mathbf{LTL}^\dagger$) rather than $\sigma_3$ alone and we follow this convention here. In this case one more operator can be built alongside the kinetic term
\begin{align}\label{CustV}
\mathcal L_{\slashed C}=\frac{v^2\varepsilon_C}{8}\textrm{Tr}\left(\mathbf{T} UD_\mu U^\dagger\right) \textrm{Tr}\left(\mathbf{T} (D^\mu U) U^\dagger\right) \,.
\end{align}
{\noindent\color{teal}{\rule{\textwidth}{1mm}}}
\textbf{Do it yourself \#2:} Custodial violation of $\mathcal L_{\slashed C}$. The argument for custodial to be a good symmetry on the bosonic sector comes from the $\rho$ parameter which is defined and measured~\cite{Workman:2022} as
$$
\rho=\frac{M_W^2}{c_{\theta_w}^2M_Z^2}=1.00038\pm 0.00040 \quad [2\sigma]\,.
$$
Translate this into a bound on $\varepsilon_C$ by finding the $W$ and $Z$ masses ($\mathcal L=1/2M_Z^2Z^2+M_W^2W^+W^-$) that follow from~(\ref{actU}) and ~(\ref{CustV}) where $$Z=c_{\theta_w}W^3-s_{\theta_w}B\,, \qquad\qquad  W^+=(W^1-iW^2)/\sqrt{2}\,.$$

{\noindent\color{teal}{\rule{\textwidth}{1mm}}}

The theory above, extended to comprise other operators and with an expansion on energy over $v$ so as to turn it into a proper EFT was outlined in 
 \cite{Appelquist:1980vg, Longhitano:1980iz,Longhitano:1980tm} where you can go for further reading.
This theory, with or without custodial, and for all its success at low energy, has a finite range of applicability and it stops at $\sim$ TeV energies. Let's see why, for simplicity neglecting $\mathcal L_{\slashed C}$ which has a coefficient bound --perhaps by yourself even-- to be small experimentally and compatible with zero.

To see this let us use the equivalence theorem~\cite{Cornwall:1974km,Vayonakis:1976vz,Lee:1977eg} which tells us that, at high energy, the amplitudes for longitudinal boson scattering $W_{L}^{\pm},Z_L$ are the same as those computed with the Goldstone Lagrangian to first order in a $M/E$ expansion with $M$ the mass of the massive gauge bosons and $E$ the energy of the scattering.

 In fact we outline here how to do this not once but twice, with two different Lagrangians. These look different yet they yield the same physics. 
 
 \begin{itemize}
 	\item [$\pi$] \textbf{Lagrangian}.
Take the CCWZ procedure and eqs.~(\ref{JVDef},\ref{GBLag},\ref{GensEW}), the kinetic term that follows from this group structure is
\begin{align}\label{piCCWZ}
-\frac{v^2}{2}(JJ)=\frac{v^2}{2}\textrm{Tr}\left(\sum_n\frac{(\left[ -i\bar\pi )^{n-1}, \partial_\mu\bar\pi\right]}{n!}X_A\right)\textrm{Tr}\left(\sum_m\frac{(\left[ -i\bar\pi )^{m-1}, \partial^\mu\bar\pi\right]}{m!}X_A\right)+\mathcal O(A_\mu)
\end{align}
where by $([ -i\bar\pi )^{n-1}, \partial_\mu\bar\pi]$ we mean $n-1$ nested commutators and we can neglect the gauge couplings since they are subleading in $M/E$.

\item[$\omega$] \textbf{Lagrangian.}
Our second Lagrangian can be built using $U$ parametrized \textit{\`a la Weinberg} (a.k.a. the square root parametrization)
\begin{align}\label{sqrtW}
	U(\omega)=i\sigma_I\frac{\omega^I}{v}+\sqrt{1-\frac{\omega^2}{v^2}}\,,
\end{align}
we find
\begin{align}\label{omegafull}
	\frac{v^2}{4}(D_\mu UD^\mu U^\dagger)=\frac12\left((\partial\omega)\cdot(\partial\omega)+\frac{(\omega\cdot\partial\omega)^2}{v^2-\omega^2}\right)+\mathcal O(A_\mu)\,.
\end{align}
\end{itemize}
As for the explicit amplitude to compute, let's focus on $W_L^+W_L^+$ scattering to leading order in our energy expansion for which purpose we define
 \begin{align}\label{PMs}
	\omega^+=&\frac{1}{\sqrt{2}}(\omega^1-i\omega^2)\,, &
	\pi^+=&\frac{1}{\sqrt{2}}(\pi^1-i\pi^2) \,,
\end{align}
and after substitution in the Lagrangian the computation 

{\noindent\color{teal}{\rule{\textwidth}{1mm}}}
\textbf{Do it yourself \#3:} Compute the amplitude by expanding the Lagrangians to second order to find, for $\pi$ and eq.~(\ref{piCCWZ})
\begin{align}\label{LagPi}
	-\frac{v^2}{2}\textrm{Tr}(JJ)=\frac{1}{2}\partial\pi\cdot\partial\pi+\frac{1}{6v^2}\left((\partial_\mu\pi\cdot\pi)^2-\pi^2(\partial\pi)^2\right)+\mathcal O(\pi^6,A\,\pi^2)\,,
\end{align}
where $\pi\cdot\partial_\mu\pi=\pi^A\partial_\mu\pi^A$ and for $\omega$, eq.~(\ref{omegafull})
\begin{align}\label{LagOm}
	\frac{v^2}{4}(D_\mu UD^\mu U^\dagger)=\frac12(\partial\omega)\cdot(\partial\omega)+\frac{(\omega\cdot\partial\omega)^2}{2v^2}+\mathcal O(\omega^6,A\,\omega^2)\,,
\end{align}
then substituting in eqs.~(\ref{PMs}) for the charged Goldstones. From there write down the Feynman rule for the 4 point vertex with external particles $\pi^+(p)\pi^+(k)\to \pi^+(p')\pi^+(k')$ and evaluate it on shell neglecting masses, $s=(p+k)^2=2pk=2p'k'$, $t=-2pp'=-2kk'$, $u=-2pk'=-2kp'$ (you can take $\partial_\mu\to -ip_\mu$ for an incoming particle and $\partial_\mu\to ip'_\mu$ for an outgoing particle recalling a $\pi^-$ field produces an outgoing $\pi^+$ particle) to find eq.~(\ref{AmpPP}).
\begin{center}
	\begin{tikzpicture}
	\draw [decorate,decoration=snake, ultra thick] (-1.732/2,1.732/2)node[anchor=east] {$W_L^+,p_\mu$} -- (0,0);
	\draw [decorate,decoration=snake, ultra thick] (-1.732/2,-1.732/2) node[anchor=east] {$W_L^+,k_\mu$} -- (0,0);
	\draw [decorate,decoration=snake, ultra thick] (0,0) -- (1.732/2,1.732/2) node[anchor=west] {$W_L^+,p'_\mu$};
	\draw [decorate,decoration=snake, ultra thick] (0,0) -- (1.732/2,-1.732/2) node[anchor=west] {$W_L^+,k'_\mu$};
	\end{tikzpicture}
\end{center}
{\noindent\color{teal}{\rule{\textwidth}{1mm}}}
returns the amplitude, our convention for the $S$-matrix being $S=1-i\mathcal A(2\pi)^4\delta^4(p)$ and $s$ the center of mass energy,
\begin{align}
	\mathcal A_{W^+_LW^+_L}=\frac{s}{v^2}+\mathcal{O}(M/E)\label{AmpPP}\,.
\end{align}
How does this amplitude tell us about the finite range of validity of our theory? A fundamental property of any theory is unitarity, which guarantees probabilities less than one,
and for the $S$ matrix translates into $SS^\dagger=1$. For ordinary matrices this would tell us we can find $S$ as an unitary matrix with say Euler's parametrization from which it follows that no single entry can be larger than one.
 The $S$ matrix space is not as simple as that but an analogous constraint can be derived and goes by the name of perturbative unitarity.
 
 {\noindent\color{teal}{\rule{\textwidth}{1mm}}}
 \textbf{Do it yourself \#4:} Derive the unitarity constraint, which can be cast as
 \begin{align}\nonumber
 	0=&(2\pi)^4\delta(p+k-p'-k')2 \,\textrm{Im} (\mathcal A)\\
 	&+(2\pi)^4\delta(p+k-q_1-q_2) \mathcal A \frac12\frac{d^3q_1d^3q_2}{(2\pi)^6(2E_{q_1})(2E_{q_2})} \mathcal A^\dagger(2\pi)^4\delta(q_1+q_2-p'-k') \,,
 \end{align}
 by performing the integral in the centre of mass frame and using $\mathcal A$'s independence of angles in our case to arrive at\footnote{Note that in some textbooks the definition $S=1+i\mathcal A (2\pi)^4\delta^4(p)$ will lead to the opposite sign for the imaginary part.
 }
 $$
 \frac{1}{16\pi}|\mathcal A(s)|^2+2\textrm{Im} (\mathcal A(s))=0\,,
 $$
 
 {\noindent\color{teal}{\rule{\textwidth}{1mm}}}
 For our amplitude perturbative unitarity  demands \begin{align}
 	\frac12\textrm{Re}\left[\int_{-1}^1 d(\cos(\theta)) \mathcal A_{W_L^+W_L^+}(s,\theta)\right]\leq 16\pi\,,
 \end{align}
with $\theta$ the scattering angle, which does not appear on eq.~(\ref{AmpPP}) but we have used the formula for generic amplitudes. Well, this is not compatible with eq.~(\ref{AmpPP}) if $E=\sqrt{s}$ goes above $\sqrt{16\pi}\,246$GeV$\sim$TeV. Unitarity, which ensures probability conservation, is not something you can give up and expect to obtain a theory that we can make sense of; it is our strongest theory argument for new physics and this theory is pointing us to TeV energies.

\section{The Higgs particle}\label{sec:Hggs}

LHC announced the discovery of the Higgs boson in 2012 to great elation and proclamations such as the last piece of the Standard Model is here.
To the theory community the discovery of the Higgs, while welcomed, was almost a given because, by the powerful unitarity arguments of the last section, we had to find \textit{some new particle} by TeV energies. The Standard Model offered the Higgs boson as that something while being in good agreement with experimental data sensitive to loop effects; the opposition, in the form of technicolour\footnote{Technicolour, in essence, replicates QCD's spontaneous breaking of the axial symmetry at the electroweak scale with new strong dynamics, see.~\cite{Lane:1993wz}}, was not so favoured by experimental data.
 
The assurance that unitarity gave us about the existence of a new particle --which, it shall be noted, is fundamentally different from  arguments deriving from the hierarchy and other fine-tuning problems-- and how the Higgs singlet can be just the right particle for the job is what we turn to next. We know that the Higgs particle, here denoted $h$, is a spin 0, parity + particle, which covers point \{\ref{ITMA}\} of our EFT list of demands, but for point \{\ref{ITMB}\} its transformation under the symmetry must be specified, (let $(h)_G$ be the Higgs field under a $G$ transformation in keeping with our notation)
$$
(h)_G=h\,.
$$
That is pretty simple, and in turns means gauge symmetry imposes no constraints on the way one can insert powers of the Higgs field anywhere. In particular if we revisit the Lagrangian of eq.~(\ref{actU}) with a Higgs we find
\begin{align}\label{acth}
\mathcal L=	\frac{(\partial h)^2}{2}+\frac{v^2F(h)^2}{4}\textrm{Tr}\left[D_\mu U D^\mu U^\dagger\right]-\frac{\mathcal P_{G}(h)}4 F_{\mu\nu}F^{\mu\nu}-V(h)-\frac{v}{\sqrt{2}}\bar\psi_L U \mathcal Y_\psi(h)\psi_R+h.c.  
\end{align}
Here we define $F(0)=1, \mathcal P_G(0)=1$ to preserve $v=246$GeV and gauge couplings $g_s,g,g'$ as usual but otherwise the only assumption about these functions is that they admint a Taylor series in $h$ around the vacuum which is $\langle h\rangle=0$ in our convention.

\subsection{Standard Model}
Before making more explicit what the electroweak EFT is, let us show why the Higgs can solve unitarity violation. Again making use of the equivalence theorem and now expanding $F^2(h)$ we have, for each $\omega$ and $\pi$ parametrization
\begin{align}\label{hGbGb}
\mathcal L_{h\pi^2}=& \frac12 2F'(0)h\, \partial \pi\cdot\partial\pi\,, &
\mathcal L_{h\omega^2}=& \frac12 2F'(0)h\, \partial \omega\cdot\partial\omega\,,
\end{align}
where we used $F(0)=1$. This adds two diagrams to $W_L^+W_L^+$ scattering, the sum of these two with the contact term 

{\noindent\color{teal}{\rule{\textwidth}{1mm}}}
\textbf{Do it yourself \#5:} Higgs contribution to longitudinal scattering. Derive the Feynman rule from eq.~(\ref{hGbGb}) after subbing in~(\ref{PMs}), don't forget the factors of $i$ and neglect the Higgs mass (including in the propagator) to add the diagrams
\begin{center}
	\begin{tikzpicture}[rotate=90]
	\draw [decorate,decoration=snake, ultra thick] (-3/2,1.732/2) -- (-1,0);
	\draw [decorate,decoration=snake, ultra thick] (-3/2,-1.732/2) -- (-1,0);
	\draw [dashed, ultra thick] (-1,0) -- (0,0);
	\draw [decorate,decoration=snake, ultra thick] (0,0) -- (1/2,1.732/2);
	\draw [decorate,decoration=snake, ultra thick] (0,0) -- (1/2,-1.732/2);
	\end{tikzpicture}
	\qquad\qquad \quad
	\begin{tikzpicture}[rotate=90]
	\draw [decorate,decoration=snake, ultra thick] (-3/2,1.732/2) -- (-1,0);
	\draw [decorate,decoration=snake, ultra thick] (1/2,-1.732) -- (-1,0);
	\draw [dashed, ultra thick] (-1,0) -- (0,0);
	\draw [decorate,decoration=snake, ultra thick] (0,0) -- (-3/2,-1.732);
	\draw [decorate,decoration=snake, ultra thick] (0,0) -- (1/2,1.732/2);
	\end{tikzpicture}
	\qquad\qquad \quad
	\begin{tikzpicture}
	\draw [decorate,decoration=snake, ultra thick] (-1.732/2,1.732/2) -- (0,0);
	\draw [decorate,decoration=snake, ultra thick] (-1.732/2,-1.732/2) -- (0,0);
	\draw [decorate,decoration=snake, ultra thick] (0,0) -- (1.732/2,1.732/2);
	\draw [decorate,decoration=snake, ultra thick] (0,0) -- (1.732/2,-1.732/2);
	\end{tikzpicture}
\end{center}
and find the amplitude in eq.~(\ref{AmpPPh}).

{\noindent\color{teal}{\rule{\textwidth}{1mm}}}
gives the amplitude, again with our approximations,
\begin{align}\label{AmpPPh}
	\mathcal A_{W_L^+W_L^+}= \left(\frac{s}{v^2}-(F'(0))^2s\right)\equiv \frac{s}{f^2}\,,
\end{align}
where we have defined
\begin{align}\label{FAmp}
	(F'(0))^2=\frac{1}{v^2}-\frac{1}{f^2}\,.
\end{align}
With our definitions we appear to be back in the same theory situation as before with the scale $f$ in place of $v$ and expecting new physics at $\sim\sqrt{16\pi}f$. Yet this is a rather different case; before the Higgs discovery physicists \textit{knew} $v=246$GeV, since it could be derived from $v=2M_W/g$, but what do we know about $f$ now? Instead of a number we can only put a lower bound on it which allows for a very large $f$, and in the limit $f\to \infty$, the amplitude vanishes to this order, and with no amplitude, no unitarity problem. That is the fundamental difference with our theory pre-Higgs discovery, the theory obtained in the limit $f\to \infty$ is perfectly self-sufficient to arbitrary energies and experimental data is compatible with this limit. This theory is none other than the Standard Model and we can define it as
\begin{align}\label{sbsSM}
	F_{SM}&=1+h/v\,, & \mathcal P_{G,SM}&=1\,, & V_{SM}(h)&=-\frac{m^2_H(v+h)^2}{2}
+\frac{\lambda(v+h)^4}{8}\,,\end{align}
\begin{align}
v\mathcal Y_q(h)&=\left(\begin{array}{cc}Y_u&\\
&Y_d\\
\end{array}\right)(v+h)\,, & 
v\mathcal Y_\ell(h)=&\left(\begin{array}{cc}0&\\
&Y_e\\
\end{array}\right)(v+h)\,.
\end{align}

From this point of view the SM does not seem such a special theory, in particular the fact that is self-sufficient, in the shape of being renormalizable, is not evident. To illustrate this property there is a re-writing in terms of a linear representation of the gauge group $H$ defined as
\begin{align}
	H=\left(\begin{array}{c}
	H^+\\
	\frac{v+H^0+i\eta}{\sqrt{2}}
	\end{array}\right)=\frac{v+h}{\sqrt{2}} U\left(\begin{array}{c}
	0\\1
	\end{array}\right)\,,
\end{align}
with $\eta$ a pseudo scalar field, $H^+$ a complex scalar field.
When written in terms of $H$ the kinetic term goes from an infinite tower of terms to
\begin{align}\label{KinH}
	\frac12(\partial h)^2+\frac{v^2}{4}F_{SM}(h)^2\textrm{Tr}\left[D_\mu U D^\mu U^\dagger\right]=D_\mu H^\dagger D^\mu H\,,
\end{align}
and Yukawa and potential read ($\tilde H=\epsilon H^*$ with $\epsilon$ the antisymmetric 2-tensor)
\begin{align}
\frac{v}{\sqrt{2}}\bar q_L U \mathcal Y_q(h)q_R&=\bar q_L\left(\tilde HY_u, H Y_d\right) q_R\,,&
	V&=-m_H^2H^\dagger H+\frac{\lambda}{2} (H^\dagger H)^2\,.
\end{align}

{\noindent\color{teal}{\rule{\textwidth}{1mm}}}
\textbf{Do it yourself \#6:} Show that relation~(\ref{KinH}) holds. You might want to use relations like $D_\mu(U^\dagger U)=0$, $D_\mu U^\dagger D^\mu U=-U^\dagger D_\mu U U^\dagger D^\mu U$ with $U^\dagger D^\mu U$ projecting only on Pauli matrices and $(0,1)^TM(0,1)=$Tr$(M(1-\sigma_3)/2)$. But feel free to use your intuition, there is more than one way to go about it.

{\noindent\color{teal}{\rule{\textwidth}{1mm}}}

It is now that we see that all the operators on this theory can be built out of linear representations with no constraints on them and have mass dimension $\leq 4$, they are marginal and relevant or renormalisable. This means that the theory is closed under quantum corrections, no counter-terms in the form of new operators will appear at any level in the theory and once we have measured all the parameters in the Lagrangian of eq.~(\ref{acth}) with the substitutions in eq.~(\ref{sbsSM}) our theory is fully predictive.

\subsection{HEFT building blocks}
It is not the SM that these notes are about so let's conclude our sightseeing trip and come back to EFT. We can summarise the two first points of the EFT list, \{\ref{ITMA}\} field and \{\ref{ITMB}\} symmetries as
\begin{align*}
 &&&q_L& &q_R&& \ell_L&& \ell_R && U && h\\\hline
&SU(3)_c&&3&& 3&& -&& -&&-&&-\\
&SU(2)_L&&2&& -&& 2&& -&& 2&&-\\
&U(1)_Y&&\frac{1}{6}&& \frac16+\frac{\sigma_3}{2}&& -\frac12&& -\frac12+\frac{\sigma_3}{2} && -\frac{\sigma_3}{2}&& -
\end{align*}
where $SU(2)_L$ ($U(1)_Y$) acts to the left (right) of $U$. If custodial is not respected we also add
\begin{align}\nonumber
 &&&SU(2)_L && U(1)_Y\\\hline\nonumber
&\mathbf{T}=U\sigma_3U^\dagger && \mathbf{3}&&-
\end{align}
where now $\mathbf{3}$ is the adjoint of $SU(2)_L$, and we note that $\mathbf{T}$ is not independent but an object built out of $U$, $\sigma_3$.

This completes the list of elements we need to build the theory which you can see in its inception on~\cite{Feruglio:1992wf,Grinstein:2007iv}.
The point that we would like to stress is that in general
 $h$ and $U$ are completely independent but in specific cases like the Standard Model they come together to form a linear representation. This is one of the main questions that these notes try to address 
 \begin{center}
 	 \textit{when and how does} HEFT \textit{admit a linear representation} $H$ \textit{?}
 \end{center}
  When HEFT does admit a linear representation we recover the Standard Model Effective Field Theory where one can write our Lagrangian in terms of
  \begin{align}\label{HlinUh}
  	H=\frac{v_\star+\underline{h}}{\sqrt{2}}\,\, U\left(\begin{array}{c}
  	0\\1
  	\end{array}\right)
  \end{align}
 where $v_\star$ and $\underline{h}$ are related but not the same as our field $h$ and constant $v$ (e.g. $\underline h$ might not be canonically normalized).
 This distinction of SMEFT and HEFT that is not SMEFT, i.e. HEFT/SMEFT, is referred to as linear vs non-linear and we will call theories in HEFT/SMEFT quotient theories to avoid the overuse of HEFT.
\begin{figure}
	\centering
	\includegraphics[width=.5\linewidth]{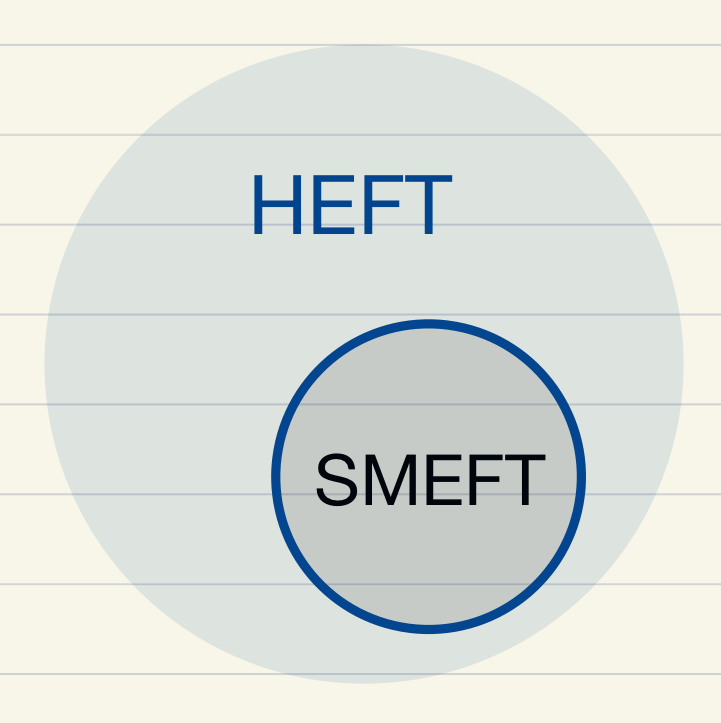}
\end{figure}

The last point in our list, \{\ref{ITMC}\}, is left to discuss, and this pertains to the selection of operators we have put in our Lagrangian thus far. Indeed in addition to the operators in eq.~(\ref{acth}) many others can be included, e.g. their square, or any positive power, as well as new combinations of our building blocks. In fact there are infinitely many operators and one needs to identify an EFT expansion to sort them out.

\section{The expansion} \label{sec:Exp}

Take our HEFT to be valid up to a scale $\Lambda$ where new particles show up to complete the picture. This variable gives us the expansion parameter since we expect our EFT to be valid at $E\ll\Lambda$ and it is a good idea to expand on powers of $E/\Lambda$. This discussion, simple as it is, is enough to pin down the organizing rule for the operators in SMEFT. It is as simple as counting mass dimensions $d_i$ of operators, and the SM is \textit{defined} to be the linear theory with all $d_i\leq 4$ operators, see e.g.~\cite{Brivio:2017vri}.

In HEFT the expansion is less evident and makes a simple counting rule like the one for SMEFT less straight forward.
Some of the difficulty is overcome by looking at amplitudes, so let's write the $W^+_LW^+_L$ amplitude with a sub-leading term in the EFT expansion,
\begin{align}\label{AmpNLO}
	\mathcal A_{W^+_LW^+_L}=\frac{s}{f^2}\left(1+c_{NLO} \frac{s}{\Lambda^2}\right)\,.
\end{align}
This equation might spark the question in the minimalist mind, given $[f]=1$ why not identify $f=\Lambda$? The two scales are different just like $v$ and $m_W$ were different in our pre-2012 theory of eq.~(\ref{actU}). Perturbative unitarity arguments give us an upper bound on $\Lambda$,  $\Lambda\leq \sqrt{16\pi}f$ which is saturated for strong dynamics, but this bound leaves a lot of room for the relation between the two scales.
 Let us be more explicit and take $\Lambda$ to be the mass and $g_*$ to be the coupling of the particle that produces $\mathcal A_{W_L^+W_L^+}$ at high energies
\begin{align}
\mathcal A_{UV}=\frac{g_*^2 s}{\Lambda^2-s}\simeq\frac{g_*^2s}{\Lambda^2}\left(1+\frac{s}{\Lambda^2}\right)\,,
\end{align}
so we can identify $\Lambda=g_*f$ and conclude that weakly coupled theories have states with particle's mass $\Lambda$ well below $f$ and strong dynamics means heavier states, with masses possibly above $f$.
 
 Turning now to the Lagrangian one shall be aware that the redundancies removed in amplitudes might obscure the view. For example, $F'$ in eq.~(\ref{FAmp}) depends on two scales but the amplitude only on one. For this reason in the following we keep the expansion implicit in $h$  within unspecified functions such as $F$ or generically $\mathcal P_i$. Similarly but now to preserve manifest gauge invariance, we do not expand $U$ on its Goldstone boson components.
 
 As a first go at organizing our operators let's mimic the SMEFT formula and write schematically a term in the Lagrangian
 \begin{align}
 \mathcal L_i=\Lambda^4\left(\frac{\psi}{\Lambda^{3/2}}\right)^{N_{\psi_i}}\left(\frac{D}{\Lambda}\right)^{N_{p_i}} (U)^{N_{U_i}}\mathcal P_i(h)\,,
 \end{align}
 where $N_i$ are natural numbers, the Lorentz and gauge indeces have been supressed and field strengths are captured as $[D_\mu,D_\nu]\sim ig F_{\mu\nu}$. Like in SMEFT we can expect higher inverse powers of $\Lambda$ to give ever less relevant corrections and operators with higher powers of fields are suppressed as intuitively expected; however upon the closer inspection of writing some of the first few terms this formula gives, we find 
 \begin{align}
 \Lambda^4 \mathcal F_V(h)+ \mathcal P_F(h) \Lambda^2  (DU DU)+\Lambda\mathcal Y(h) \psi^2 U+\mathcal P_{4}(h)(DU)^4+\cdots\,,
 \end{align}
 which suggests the potential is much more relevant than the kinetic term of the Goldstones and this more so than the Yukawa term, itself less important that kinetic terms of fermions and gauge bosons, the latter on the same footing as the last operator above. The last operator above, $(DU)^4$, however gives the sub-leading correction in eq.~(\ref{AmpNLO}) which would require
 $\mathcal P_4(0)\sim v^4/(f^2\Lambda^2)$ and hence a dependence on $\Lambda$ that the SMEFT formula failed to predict. 
 
 A more appropriate counting formula to organise our EFT can be obtained from Naive Dimensional Analysis (NDA)~\cite{Manohar:1983md}, to which we turn next. NDA gives a normalization of operators such that the dimensionless coefficients that accompany them in the Lagrangian should be smaller than one for the loop expansion to converge.
 
{\noindent\color{teal}{\rule{\textwidth}{1mm}}}
\textbf{Do it yourself \#7:} Understanding naive dimensional analysis.
 
 To implement the NDA normalization based on the loop expansion one can proceed as follows: we first divide field insertions by a scale, let's call it $\Lambda_\phi$ for scalars and impose that in the
  propagation of our field an inverse loop factor arises together with a common scale $\Lambda$
 \begin{align}
 \langle 0|\frac{\phi(x) \phi(y)}{\Lambda_\phi^2}|0\rangle=\int \frac{d^4p}{(2\pi)^4}\frac{ie^{ip(x-y)}}{p^2-m^2}\frac{1}{\Lambda_\phi^2}\sim (4\pi)^2\int\frac{e^{ip(x-y)}d^4p}{(2\pi)^4(p^2-m^2)\Lambda^2},
 \end{align}
 so $\Lambda_\phi=\Lambda/(4\pi)$, you can do the same for fermions to find $\Lambda_f^3=\Lambda^3/(4\pi)^2$. Powers of momenta are inserted over this scale $\Lambda$.
 
 Next consider an operator in the Lagrangian with the normalisation above --and an overall $\Lambda^4/(4\pi)^2$-- and $E_\phi$ scalars --we leave other fields out for simplicity-- and $K$ derivatives with a coefficient $C$ at tree level. At the quantum level this operator will receive a contribution $\delta C$ from a diagram with $V$ vertexes, each contributing $E_{\phi_i}$ external fields, $N_i$ powers of momenta and a total of $I$ internal lines, in momentum representation,
 \begin{align}
 &\delta C\frac{\Lambda^4}{(4\pi)^2}\left[\frac{4\pi\phi}{\Lambda}\right]^{E_\phi}
 \left(\frac{p}{\Lambda}\right)^K(2\pi)^4\delta^4(p)\\\nonumber
 &=\prod_i^V\left(\frac{\Lambda^4}{(4\pi)^2}\left[\frac{4\pi\phi}{\Lambda}\right]^{E_{\phi_i}}\left(\frac{p}{\Lambda}\right)^{N_i}
 (2\pi)^4\delta^4(p)\right)\prod_j^I\left(\frac{dp^4_j (4\pi)^2}{(2\pi)^4p^2\Lambda^2}\right)\,.
 \end{align}
 If we use the Dirac delta's to cancel out internal momenta we are left with $L=I-(V-1)$ integrals over momenta (if you're not familiar with it, I encourage you to test the relation $L=I-(V-1)$ on the most complicated loop diagram you can scribble) to simplify the contribution to $C$ as ($N=\sum N_i$)
 \begin{align}
 \delta C\sim &(\prod C_i)\left(\frac{\Lambda^4}{(4\pi)^2}\right)^{V-1}\left(\frac{\ell}{\Lambda}\right)^{N-K}\left[\frac{d^4\ell}{(2\pi)^4}\right]^L\left[\frac{(4\pi)^2}{\ell^2\Lambda^2}\right]^I\\
 \sim& (\prod C_i)(\Lambda^4)^{V-1+L-I}(4\pi)^{I-L-V+1}=(\prod C_i)\,,
 \end{align}
 which means the contribution to $C$ is the product of all coefficients with no $\pi$-factor suppression and $C_i\sim1$ is the limit of convergence of our loop expansion, anything larger than that makes computations unviable.
 
 {\noindent\color{teal}{\rule{\textwidth}{1mm}}}
 
 Given our reasons not to expand on the Higgs singlet and Goldstones, the NDA adaptation reads~\cite{Gavela:2016bzc}
 \begin{align}
 \mathcal L_i=\frac{\Lambda^4}{(4\pi)^2}\mathcal P_i(h)\left[\frac{4\pi\psi}{\Lambda^{3/2}}\right]^{N_{\psi_i}}\left[\frac{D}{\Lambda}\right]^{N_{p_i}} U^{N_{U_i}}\,,
 \end{align}
 now with this formula the four derivative operator 
\begin{align}\label{OpCS1}
 \frac{\mathcal P_4(h)}{(4\pi)^2}(DU)^4\,,
 \end{align}
 is subleading and we are closer to natural coefficients with $\mathcal P_4(0)\sim v^4(4\pi)^2/f^2\Lambda^2$ to reproduce eq.~(\ref{AmpNLO}). NDA offers chiral counting $N_\chi=N_{p_i}+N_{\psi_i}/2$ to organize our series in a manner more informed than simple dimensional analysis~\cite{Gavela:2016bzc}. It does also signal the following operator  potentially as important as gauge bosons couplings to fermions 
 \begin{align}\label{OpCS2}
 \mathcal P_{\psi^2D}(h)\bar\psi_L\gamma_\mu \sigma_I\psi_L\,\textrm{Tr}\left(\sigma^I(D_\mu U)U^\dagger\right)\,.
 \end{align}
 What the NDA-adapted formula does not however make explicit is how the limit $\Lambda\to\infty$ returns the SM. Take the operators in eq.~(\ref{OpCS1}), and (\ref{OpCS2}); their coefficients
 $\mathcal P_{\psi^2D}(0)$ and $\mathcal P_{4}(0)$ have zero mass dimension so one takes them as coupling constants independent of each other and $\Lambda$ --the NDA formula does not know about eq.~(\ref{AmpNLO}); however, we know they should vanish in the limit $\Lambda\to\infty$.
  There is therefore further implicit dependence on the dimensionless coefficients $\mathcal P(0)$ that NDA cannot determine.
 
 An approach to the expansion that makes the decoupling more evident is to define your leading order Lagrangian and then deduce your next-to-leading by identifying the terms required by renormalization at one loop. In essence it uses the loop expansion as NDA does but there is the somewhat arbitrary definition of leading order (LO). Following \cite{Buchalla:2013eza} in this ordering the LO Lagrangian is defined as
 \begin{align}
 \mathcal L_{\textrm{LO}}=
 \frac12(\partial h)^2+\frac{v^2}{4}F(h)^2\textrm{Tr}\left[D_\mu U D^\mu U^\dagger\right]-\frac{1}4 F_{\mu\nu}F^{\mu\nu}-V(h)-\bar\psi_L U \mathcal Y(h)\psi_R+h.c. 	
 \end{align} 
 where we note the Higgs function $\mathcal P$  has been dropped in the gauge kinetic term. The NLO terms can be derived from induced UV divergences as, again schematically,
 \begin{align}
 \psi^2DU\,,& & \psi_L\psi_R U&[D,D]\,, &&D^4U,h \,,&&\psi^4\,.
 \end{align}
 More than estimating has been done nonetheless, the UV divergences and RGE equations have been computed in~\cite{Alonso:2017tdy,Buchalla:2020kdh}.
 
 Once the expansion has been determined, the operators have to be made explicit and redundancies removed --unless you are okay with working with more parameters than physically observable which is sometimes of use. In these notes we do not aim at presenting the complete lists of subleading operators in each framework but refer the reader to the literature \cite{Alonso:2012px,Contino:2013kra,Buchalla:2013rka,Gavela:2016bzc,Graf:2022rco}.
 
Before moving on, a word of encouragement if you're put off by the expansion ambiguity; we have experiment to tells us what the coefficients are! 

\section{Geometry}\label{sec:Geo}

Amplitudes remove redundancies from the Lagrangian description and make our expansions clearer. We have seen an example of each in $(i)$ the same amplitude following from Lagrangians~(\ref{LagPi}) and~(\ref{LagOm}), and $(ii)$ how eq.~(\ref{AmpNLO}) helps us identify the EFT expansion. Point $(i)$ can indeed be presented by saying that the physics does not care about the way we choose to parametrize field-space. This same statement applies in an analogue that might come to mind; the physics are the same no matter how we parametrize (choose coordinates of) space-time. This parallel between field reparametrisation in QFT and coordinate changes in general relativity goes further than a sketch; amplitudes are built out of tensors in field-space and we will present a `geometrised' relation between amplitudes and correlation functions. Last but not least in our list of reasons to use geometry, is that it helps us visualize our theories. Further reading on this quickly developing topic can be found in references ~\cite{Alonso:2015fsp,Alonso:2016oah,Cohen:2020xca,Cohen:2021ucp,Alonso:2022ffe}. 

How does geometry come about? It starts with a metric, and we have been dealing with a couple already, let us write eq.~(\ref{omegafull}) explicitly:
\begin{align}
	\mathcal L=\frac12 \partial_\mu\omega^A\left[\delta_{AB}+\frac{\omega_A\omega_B}{v^2-\omega^2}\right]\partial^\mu\omega^B\equiv \frac12 \partial_\mu\omega^AG^{(\omega)}_{AB}\partial^\mu\omega^B\,.
\end{align}
One can regard $G^{(\omega)}_{AB}$, a two index object, as a candidate but, for it to be a true metric, it should transform covariantly. Assume we are given the mapping from $\pi$ to $\omega$, $\omega(\pi)$, then we can write 
\begin{align}
\frac12 \partial_\mu\omega^AG^{(\omega)}_{AB}\partial^\mu\omega^B=
\frac12 \partial_\mu\pi^C\frac{\partial\omega^A}{\partial\pi^C}G^{(\omega)}_{AB}\frac{\partial\omega^B}{\partial\pi^D}\partial^\mu\pi^D=\frac12\partial_\mu \pi^C G^{(\pi)}_{CD}\partial^\mu \pi^D\,,
\end{align}
$G^{(\omega)}$ does transform covariantly and is indeed a metric. The matrices $\partial\omega/\partial \pi$ are the equivalent of the transformation $G$ in our gauge theory and 
we have a relation between the new and old metric which is of the covariant form. 

This is telling us more explicitly that both theories are the same, and for this it is useful to take a step back and realize that both parametrise differently  $2\times 2$ unitary matrices, i.e.
$$
U^\dagger U=UU^\dagger=1\,,\qquad\qquad\qquad \det(U)=1\,,
$$
but these are none other than the equations for a 3-sphere, $S^3$.

 Having identified our metric let us turn to derivatives. In an analogy with gauge theory, there is the need to modify derivatives to make them covariant, only now derivatives with respect to the field. Consider $V(\phi)$ with $\phi^a$ our set of coordinates and a change to a new set of coordinates $\tilde\phi$, $\phi(\tilde\phi)$
\begin{align}
	\frac{\partial V}{\partial \phi^a}&=\frac{\partial\tilde\phi^c}{\partial \phi^a}\frac{\partial V}{\partial \tilde\phi^c}\,,
	&
	\frac{\partial^2 V}{\partial\phi^b\partial \phi^a}&=\frac{\partial^2\tilde\phi^c}{\partial \phi^b\partial \phi^a}\frac{\partial V}{\partial \tilde\phi^c}+\frac{\partial\tilde\phi^d}{\partial\phi^b}\frac{\partial\tilde\phi^c}{\partial\phi^a}\frac{\partial^2V}{\partial\tilde\phi^c\partial\tilde\phi^d}\,.
\end{align}
While the first terms gives a covariant relation between old and new derivatives with respect to the fields, the second one has a piece that does not follow this rule. To remedy this we introduce the Christoffel symbols or connection, denoting the metric in $\phi$ coordinates $G^{(\phi)}$, 
\begin{align}\label{Conn}
\Gamma^i_{jk}=\frac{1}{2}(G^{(\phi)})^{il}\left(\partial_{j}G_{kl}^{(\phi)}+\partial_{k}G_{lj}^{(\phi)}-\partial_{l}G_{jk}^{(\phi)}\right)\,,
\end{align}
and introduce the covariant-field derivative $\mathcal D$ which acts as:
\begin{align}
\mathcal D_a \mathcal D_b V= \frac{\partial^2V}{\partial\phi^a\partial\phi^a}-\Gamma^c_{ab}\frac{\partial V}{\partial\phi^c}\,,
\end{align}
with now $\mathcal D^2V$ being a true tensor, i.e. transforming covariantly with a $\partial\phi/\partial\tilde\phi$ matrix on each index.

This analysis on HEFT gives us the metric,
\begin{align}\label{Gphi}
\frac{(\partial_\mu h)^2}{2}+\frac{(vF)^2}{4}\textrm{Tr}\!\left(D^\mu U^\dagger D_\mu U\right)&=\frac12D_\mu \phi^iG_{ij}^{(\phi)}(\phi)D^\mu\phi^j\,, & G_{ij}^{(\phi)}=&\left(\begin{array} {cc}
1&0\\
0&F^2(h)\hat g_{ab}
\end{array}\right),
\end{align}
where $\phi=(h,\varphi^a)$, $D_\mu\phi=(\partial_\mu h, \partial_\mu \varphi^a+g A_\mu^B \zeta^a_B)$ with $\zeta$ killing vectors, $\hat g$ is the metric of a 3-sphere $S^3$, $A_\mu^B=$Tr$(T_B A_\mu)$,
and $i$ runs on the values $i=h,1,2,3$. An explicit inner metric $\hat g$ in Goldstone space could be $G^{(\omega)}$ or $G^{(\pi)}$ identiying $\varphi$ with the respective coordinates but one need not and we will not specify one, noting nonetheless that our convention around the vacuum is $\hat g(0)_{ab}=\delta_{ab}$. 

The Killing vectors $\zeta$ give us the isometries of the manifold, i.e. displacements that leave the metric the same, and form a representation of the gauge group as
\begin{align}\label{KillProp}
\frac{\partial\zeta^k}{\partial \phi^i}G^{(\phi)}_{kj}
+G^{(\phi)}_{ik}\frac{\partial\zeta^k}{\partial \phi^j}+\zeta^k\frac{\partial G_{ij}^{(\phi)}}{\partial\phi^k}=0\,,\qquad\qquad \zeta^k_A\frac{\partial \zeta^i_B}{\partial\phi^k}
- \zeta^k_B\frac{\partial \zeta^i_A}{\partial\phi^k}=f_{AB}{}^{C}\zeta^i_C\,,
\end{align}
where $B$ on $\zeta^i_B$ runs over the generators of the gauge symmetry.  Killing vectors offer the answer to the question we posed in sec.~\ref{sec:Long}, eq.~(\ref{XiTExp}); the infinitesimal transformation of our fields is proportional to the Killing vectors $$\delta_G\phi^i=\epsilon^A\zeta^i_A(\phi)\,,$$ whereas the derivative, making use of the second property in eq.~(\ref{KillProp}) 
\begin{align}
	\delta_G (D_\mu \phi^i)= (D_\mu \phi^j)\frac{\partial \zeta^i_A}{\partial \phi^j} \epsilon_A  \,,
\end{align}
{\color{teal} ${||}$ which you can derive yourself:} use (\ref{GaugG}), (\ref{Lie}) and (\ref{KillProp}) {\color{teal} $||$}
and the first property of (\ref{KillProp}) shows that the kinetic term in (\ref{Gphi}) is indeed invariant. One can hence regard the Goldstone boson study of sec~\ref{sec:Long} as the application of geometry to the coset spaces that appear in group theory.

It might seem that we have introduced quite a few new concepts for a bit of a dry section so far, so let's pause here for a respite to pick some low hanging fruit in geometry. Our scalar fields, Higgs$+$Goldstones, live in a 4-dimensional manifold $\mathcal M$ with metric $G^{(\phi)}$. What could this space be?

There is a lot of possibilities of course but if one surveys for simplicity only the \textit{most symmetric} spaces for the manifold $\mathcal M$, they happen to have constant curvature $R$ over all the manifold and you might have already guessed a couple of them; here is the three we have:
\begin{align}
\textrm{most symmetric } \mathcal M's =\left\{\begin{array}{ccc}
R=0&I\!\!R^4 & \textrm{Standard Model}\\
R>0&S^4& \textrm{Minimal Composite Higgs}\\
R<0&\mathcal{H}^4& \textrm{???}
\end{array}\right.
\end{align}

\begin{figure}[h!]
	\centering
	\includegraphics[width=0.5\textwidth]{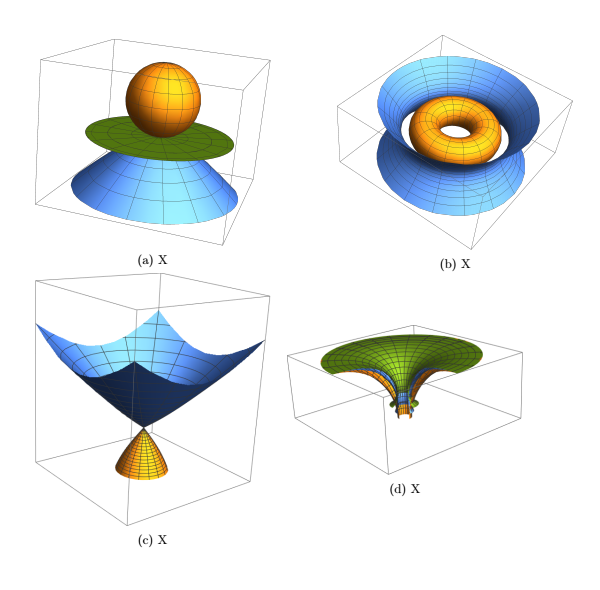}
	\caption{Examples of possible HEFTs based on geometry and reduced to 2d (rotations around the vertical axis correspond to gauge transformations): (a) SM (green) MCHM (orange) and negative curvature (blue), (b) Quotient theories which do not have fixed point (the manifold does not go through the origin), (c) Quotient theories with a singularity at the fixed point (d) theories as in (b) but resembling ever more the SM case around $h=0$. See~\cite{Alonso:2021rac} for more details.}
\end{figure}
The last column is given without a proof, but a plausibility argument for the Minimal Composite Higgs model is that it is based on the breaking $SO(5)/SO(4)$ and this space that yields the Higgs doublet is isomorphic to a 4-sphere, $S^4$.
The $\mathcal H^4$ case has negative curvature and it might be described arising in $SO(4,1)/SO(4)$ breaking but its formulation as a complete theory runs into ghosts~\cite{Falkowski:2012vh,Alonso:2016btr}.

{\noindent\color{teal}{\rule{\textwidth}{1mm}}}
\textbf{Do it yourself \#8:} Derive the metric for the MCHM. The minimal composite Higgs model is based on the breaking $SO(5)/SO(4)$ which returns 4 Goldstones that the Higgs doublet arranges into. We can use the CCWZ formalism of sec.~\ref{sec:Long} with matrices
\begin{align}
X_A&=-i\frac{1}{\sqrt{2}}\left(\begin{array}{cc}
0&\delta_{iA}\\
-\delta_{iA}&0
\end{array}\right)\,, & i\bar\pi\cdot X=\frac{1}{\sqrt{2}f}\left(\begin{array}{cc}
0&\vec{\pi}\\
-\vec{\pi}&0
\end{array}\right)\,,
\end{align}
where we have substituted $v\to f$ and $\pi^A$, $A=1,2,3,4$.
An explicit form of $\xi$ can be found by first deriving the relations
\begin{align}
\left(i\bar\pi \cdot X\right)^{2n}=&\left(\frac{-\pi^2}{2f^2}\right)^{n}\left(\begin{array}{cc}
1&0\\
0& \frac{\pi_i\pi_j}{\pi^2}
\end{array}\right)\,,&
\left(i\bar\pi \cdot X\right)^{2n+1}=&\left(\frac{-\pi^2}{2f^2}\right)^{n}i\bar \pi \cdot X\,,
\end{align} and then applying them in the exponential expansion of $\xi$. It is  useful to define $\vec\pi^2=\underline h^2$ and $u\equiv\vec{\pi}/|\pi|$ so $u(\varphi)$ is a unit-modulus 4d vector and $\partial_\mu\varphi^a\hat g_{ab}\partial^\mu\varphi^b=v^2(\partial_\mu u)\cdot(\partial^\mu u)$. Once the form of $\xi$ is found, compute $J$ as in~(\ref{JVDef}) and from there the Lagrangian from~(\ref{GBLag}). The last step for the connection is to write $\underline h=v_\star+h$. You can find more such decompositions in~\cite{Alonso:2014wta}. For those willing to go further, the relation between $v,v_\star$ and $f$ can be found and substituted back to find $F(h;v,f)$.

{\noindent\color{teal}{\rule{\textwidth}{1mm}}}

Let us keep the geometry options open nevertheless and now push ahead to relate geometry and amplitudes. For this purpose we compute the Riemann tensor that follows from the metric $G^{(\phi)}$.

{\noindent\color{teal}{\rule{\textwidth}{1mm}}}
\textbf{Do it yourself \#9:} Derive the connection from the metric in ~(\ref{Gphi}) using formula~(\ref{Conn}), and from there the Riemann tensor given
\begin{align}
R^i_{\,\,jkl}=\partial_k \Gamma^{i}_{jl}+\Gamma^{i}_{km}\Gamma^{m}_{jl}-\left(k\leftrightarrow l\right)\,.
\end{align}
The pure Goldstone elements are a bit involved, you can try and use an explicit inner metric or relations like $\hat g_{ab}=v^2(\partial u^i(\varphi)/\partial\varphi^a)\delta_{ij}(\partial u^j(\varphi)/\partial\varphi^b)$ with $u$ a 4d unit vector.

{\noindent\color{teal}{\rule{\textwidth}{1mm}}}

The Christoffel symbols read
\begin{align}
\Gamma_{ab}^h&=-F'F\hat g_{ab}\,, & \Gamma_{hb}^a&=\frac{F'}{F}\delta^a_{b}\,,&  \Gamma_{bc}^a&=\hat\Gamma^a_{bc}\,,
\end{align}
where we need not specify the inner 3-sphere connection $\hat\Gamma$. 
The Christoffel symbols are not covariant themselves so will not feature in amplitudes; note that they show up in the Lagrangian though as evidence of redundancies.
The curvature on the other hand is covariant and reads
\begin{align}
 R_{habh}&=-F^2 R_h\hat g_{ab}\,,\\
 R_{abcd}&= F^4 R_\varphi (\hat g_{ac}\hat g_{bd}-\hat g_{ad}\hat g_{bc})\,,
\end{align}
with
\begin{align}\label{Rdef}
 R_h&\equiv-\frac{F''}{F}\,, &  R_\varphi&\equiv\frac{1}{v^2F^2}-\frac{(F')^2}{F^2}\,, &R^{ij}_{\,\,\,\,\,ij}=6(R_\varphi+R_h)\,,&&
\end{align}
you might notice, when evaluated at the vacuum where $F(0)=1$, that the curvature $R_\varphi$ features in the amplitude~(\ref{AmpPPh}), what is more one can also find
\begin{align}
	\mathcal A_{W^+_LW^-_L\to h h}=-R_h s\,.
\end{align}
 Are these coincidences?

\subsection{Geometric LSZ reduction formula}

Let us revise the LSZ reduction formula
\begin{align}
	S_{\textrm{in}\to\textrm{ out}}=\left(\left(\prod_i^n\frac{p^2_i-m^2_i}{\sqrt{Z}}\frac{\delta}{i\delta J_{a_i}}\right) \frac{\int [d\phi] e^{iS[\phi]+J\cdot\phi}}{\int [d\phi] e^{iS[\phi]}}\right|_{J=0}=\left(\prod_i^n\frac{p^2_i-m^2_i}{\sqrt{Z}}\right)\frac{\langle\phi^{a_1}\phi^{a_2}\cdots \phi^{a_n}\rangle}{\langle\,\,\rangle}
\end{align}
where $\langle\phi^{a_1}\phi^{a_2}\cdots \phi^{a_n}\rangle$ is the n-point correlation function and the kinetic term is not canonically normalized; it reads $Z^{-1}\partial\phi^2/2$ so that $\langle\phi\phi\rangle=iZ/(p^2-m^2)$.
Try now to compare this with our geometry. 

First we expect the correlation function, given it has $n$-field-indexes to generalize to a tensor. Coordinates $\phi$ are not themselves covariant; to fix this one can exchange them for a parametrization with new variables $\eta$ that follow geodesics (Riemann Normal Coordinates) and yields the same $S$ matrix elements since $\phi=\eta+\mathcal O(\eta^2)$ and $\eta$ excite the same particle as $\phi$ out of the vacuum giving the same scattering matrix $S$.

The other element is the normalization $Z$. If un-normalized, our kinetic terms in the geometric picture read $G(0)(\partial\phi)^2/2$ hence $Z^{-1}=G$ and $(Z)^{-1/2}=\sqrt{G}$; well, what is the square root of a metric? Recalling our General Relativity modules
\begin{align}\label{vierbein}
G_{ij}^{(\phi)}\equiv \sum_{IJ} \mathbf{e}_i^I(\phi)\mathbf{e}_j^J(\phi)\delta_{IJ}
\end{align}
$\mathbf{e}_i^I$ are the vierbeins which map between `flat space' ($I,J$ indexes) with a metric $\delta_{IJ}$ to our curved space of metric $G$.

The geometric generalization of the LSZ formula is then --this formula was first presented in~\cite{Cheung:2021yog}--
\begin{align}\label{LSZAW}
S_{\textrm{in}\to\textrm{ out}}=\left(\prod_i^n(p^2_i-m^2_i)\mathbf{e}_{a_i}^{J_i}\right)\frac{\langle\phi^{a_1}\phi^{a_2}\cdots \phi^{a_n}\rangle}{\langle\,\,\rangle}
	\equiv \left(\prod_i^n\mathbf{e}_{a_i}^{J_i}\right)\mathcal T^{a_1\dots a_n}(\{p_i\})\,,
\end{align} 
with $\mathcal T$ a tensor with $n$ indexes that also has kinematic variable dependence. Note that in the flat space the vierbeins map into the metric is the identity and hence invariant under a rotation which can be used to diagonalise the mass matrix.

{\noindent\color{teal}{\rule{\textwidth}{1mm}}}
\textbf{Do it yourself \#10:}
Well, don't believe me, let's try it! Suppose the vev of $h$ is not at 0 but some other point $h_\star$. 

First we can find the LHS of LSZ computing the amplitude as we did in for~(\ref{AmpPPh}), now expanding around $h_\star$
$$
\frac{v^2F(h_\star)^2}{4}\textrm{Tr}\left(D_\mu U(\omega/v_\star)D^\mu U^\dagger(\omega/v_\star)\right)+\frac{v^2F(h_\star)F'(h_\star)h}{2}\textrm{Tr}\left(D_\mu U(\omega/v_\star)D^\mu U^\dagger(\omega/v_\star)\right)
$$
where $U(\omega/v_\star)$ is the parametrisation of eq.~(\ref{sqrtW}) with $v\to v_\star$; the first term here will give us $v_\star$ in terms of $v$ and $F(v_\star)$ when we canonically normalize $\omega$, then we substitute this in the $\omega^2h$ and $\omega^4$ vertexes and the new coefficient with our old computation should give us the new amplitude. 

On the RHS of the equation
we would have, for this scattering, given that $\omega^+$ is a combination of $\omega^{1,2}$,  
$$\mathcal A_{W_L^+W_L^+}=(\mathbf{e}_{i}^1(h_\star))(\mathbf{e}_{j}^2(h_\star))(\mathbf{e}_{k}^1(h_\star))(\mathbf{e}_{l}^2(h_\star))R^{ijkl}(h_\star)s$$
  We evaluate our equation at $h=h_\star$ but still around the same point in the Goldstone manifold, in particular $\hat g(0)_{ab}=\delta_{ab}$; use this and eqs.~(\ref{Gphi},\ref{vierbein}) to find the explicit form of $\mathbf{e}_i^{I}$ and evaluate the equation above to check eq.~(\ref{LSZAW}).

{\noindent\color{teal}{\rule{\textwidth}{1mm}}}

One can check both sides of this equation with the amplitude at hand, if the Higgs boson happens to take a vev at some other $h_{\star}$ recycling our result with the RHS of the geometrised LSZ formula, we would find
$$\mathcal A_{W_L^+W_L^+}= F(h_\star)^4 R^{1212}(h_\star) s=R_\varphi(h_\star) s\,,
$$
as one can check explicitly computing the amplitude.

\subsection{The metric for the linear realisation}

Geometry offers therefore a good grasp on physical observables. Let's put it to use to come back to the question we posed in sec.~\ref{sec:Hggs}, the distinction between linear (a.k.a. SMEFT) and non-linear (a.k.a. quotient theories, HEFT/SMEFT). 

For a first look lets assume SMEFT with a good expansion so it suffices to consider the leading corrections; with the extra assumption of custodial invariance, the two operators in SMEFT that contribute to the metric are
\begin{align}
\mathcal L_{d=6}=	\frac{c_{H\Box}}{2\Lambda^2}(\partial_\mu (H^\dagger H))^2+\frac{c_{HDD}}{\Lambda^2}H^\dagger H D_\mu H^\dagger D^\mu H\,.
\end{align}
These come with two parameters so one might expect physical observable predictions that depend on two variables. This is not the case.

{\noindent\color{teal}{\rule{\textwidth}{1mm}}}
\textbf{Do it yourself \#11:} Check that leading order SMEFT predicts correlated curvature. There is a number of ways of checking this: computing the amplitudes that follow from the Lagragian, using the equations of motion, and others; what we can try here is more geometric. 
After substituting $(\ref{HlinUh})$ in the operators (and Higgs doublet kinetic term) the metric has a form 
\begin{align*}
D_\mu H^\dagger D_\mu H+\frac{c_{H\Box}}{2\Lambda^2}(\partial_\mu (H^\dagger H))^2+\frac{c_{HDD}}{\Lambda^2}H^\dagger H D_\mu H^\dagger D^\mu H
&\equiv \frac{K(\underline{h})(\partial\underline h)^2}{2}+ \frac{L^2(\underline{h})}{2}(d\Omega^2)\\
&=\frac{(\partial h)^2}{2}+ \frac{v^2F^2(h)}{2}(d\Omega^2)
\end{align*}
where $d\Omega^2=$Tr$(D_\mu UD^\mu U^\dagger)/2$. The second equality gives us the change of coordinates from which you can derive the relations, using the chain rule and eq.~(\ref{Rdef}), 
\begin{align}\label{RDefNew0}\frac{d\underline h}{dh}=\frac{1}{\sqrt{K}}, \qquad R_\varphi=\frac{1}{L^2}-\frac{1}{KL^2}\left(\frac{dL}{d\underline h}\right)^2,\qquad R_h=-\frac{1}{KL}\frac{d^2L}{d\underline h^2}+\frac{1}{2LK^2}\frac{dL}{d\underline h}\frac{dK}{d\underline h}.
\end{align}
 Whereas the first identity gives us $K(\underline h),L(\underline h)$ in terms of the operators, which we can substitute above and expand all terms to order $\Lambda^{-2}$ to verify eq.~(\ref{Corr}).

{\noindent\color{teal}{\rule{\textwidth}{1mm}}}
as one can derive the result that
\begin{align}\label{Corr}
	R_\varphi=R_h=\frac{c_{H\Box}-c_{HDD}}{\Lambda^2}\,,
\end{align}
This is a definite prediction that we can compare with data by looking for correlations between $W_LW_L\to W_LW_L$ and $W_LW_L\to hh$ amplitudes (in practice however these are hard observables to observe and we probe curvature through other means, see~\ref{sec:PH}. However, say we find experimentally --which we haven't-- $R_h=R_\varphi/3$; this does not agree with (\ref{Corr}), can we rule out SMEFT?
 
The answer is no; if one does not stop at dimension 6 but adds further terms
\begin{align}
	\frac{1}2(\partial h)^2+\frac{v^2F^2}{2}d\Omega^2 = A\left(H^\dagger H\right)(\partial_\mu (H^\dagger H))^2+B\left(H^\dagger H\right) D_\mu H^\dagger D^\mu H\,,
\end{align}
where $d\Omega^2=$Tr$(D_\mu UD^\mu U^\dagger)/2$, with $A,B$ analytic functions of the argument, the correlation is lost; it already is at dimension 8 actually. There has to be a better way of telling SMEFT and HEFT/SMEFT apart.

\section{The EFT Dichotomy}\label{sec:Dic}

The way to tell linear and non-linear realisations was laid out by the CCWZ authors in the same work~\cite{Coleman:1969sm} quoted in sec~\ref{sec:Long}, and cast into HEFT in~\cite{Alonso:2016oah,Cohen:2020xca}. To present it we point out first the implicit assumption in SMEFT that an expansion in $H$ is well behaved, i.e. analytic in at least a neighbourhood of $H=0$.
Yet this point is not the vacuum state, since  $\langle H^\dagger H\rangle=v_\star^2/2$; in contrast for HEFT one expands on $h$ where $h$ excites the physical Higgs and parametrises deviations of the field from the vacuum point. 

What is special about $H=0$, why would SMEFT expand around it rather than the vacuum? Electroweak symmetry is restored, i.e. it is unbroken at the point $H=0$. The presence or not of such point in our manifold is the key to the lemma to tell SMEFT and HEFT/SMEFT apart\\

\textit{ Linearisation lemma} 
\begin{center}
 If there exists a fixed point in $\mathcal M$ under the $\mathcal G$ group action
with a smooth neighbourhood then the theory admits a linear representation i.e.
$$\exists h_\star \quad \textrm{such that}\quad (i) \,\,\,F(h_\star)=0 \quad\textrm{\&} \quad(ii) \,\,\, (\nabla^n R)\leq \infty\,\,\forall\, \,n\quad \Rightarrow\quad \mathbf{Linear}$$
\end{center}
A fixed point is defined to be left unchanged under the action of the group, fixed; for the Higgs doublet
$$
(H)_G=GH=H?\qquad  \qquad\textrm{ only solution }\quad H=0 \,.
$$ 
The second line of the lemma phrases it for our metric of~(\ref{Gphi}); Goldstones are always shifted by $G$, so the only way to avoid this is by `collapsing' the angle coordinates to a point by setting $F(h_\star)=0$ just as it happens at the origin for polar coordinates.
There you have it then, if a smooth fixed point exists, i.e. conditions $(i,ii)$ are met, we have SMEFT.

It also tells us that breaking either of these conditions one can step out of SMEFT and the linear realisation into quotient theories, which can therefore be put in two classes
\begin{itemize}
	\item [\textbf{A}] There does not exist a point $h_\star$ left fixed by the group action   ($\nexists h_\star\,,\,F(h_\star)=0$)
	\item [\textbf{B}] There is a fixed point $h_\star$ with $F(h_\star)=0$ but is singular, i.e. $\nabla^nR=\infty$ for some $n$
\end{itemize}

Before we see one of these quotient theories is good to review the familiar theories for BSM, to see if any of them fits the mould. Supersymmetry, Grand Unified Theories, the Seesaw model and even composite Higgs models, all fall in the SMEFT category. They also have one thing in common, their typical scale can be much higher than the electroweak scale and hence evade direct searches. 
 This connects to an important theorem in particle physics. The decoupling theorem~\cite{Appelquist:1974tg} guarantees that if the new physics behind our EFT is heavy enough, it would decouple to leave no effects and a consistent theory. This is the case in SMEFT as taking $\Lambda\to\infty$ we obtain the Standard Model. However is this the only way to decouple our theory? In other words, are HEFT/SMEFT quotient theories non-decoupling which would imply they have a finite cut-off --here meaning by cut-off the mass of the extra heavy states? Recent study suggest this is the case, at least for a large class of quotient theories~\cite{Falkowski:2019tft,Cohen:2021ucp,Alonso:2021rac}; the opposite would mean there is a `back door' to the SM via HEFT/SMEFT.
\begin{figure}[h]
	\centering
	\includegraphics[width=.75\linewidth]{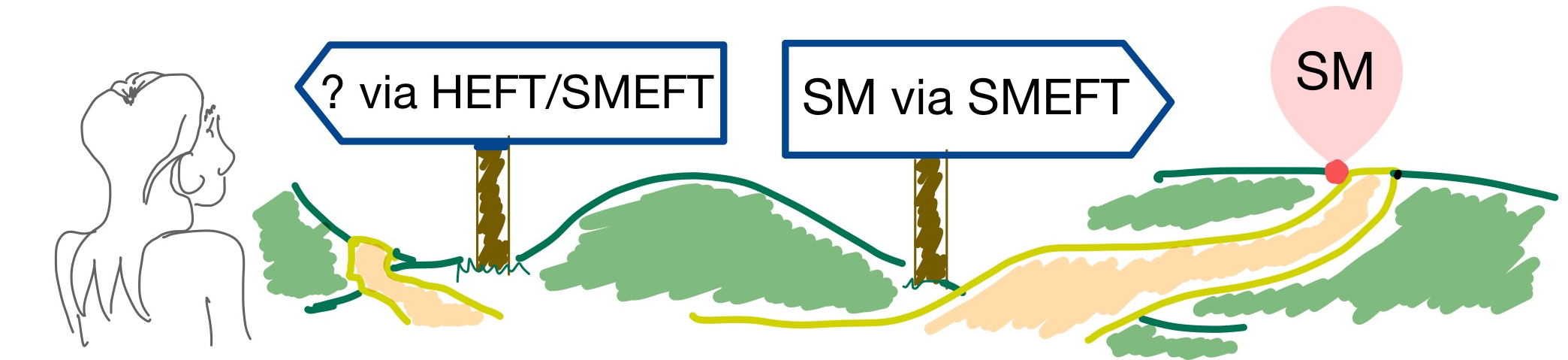}
\end{figure}

For pedagogical reasons, rather than reviewing the general arguments or proofs for the properties of quotient theories, let us simply build a specific theory that interpolates between SMEFT and HEFT/SMEFT and examine it. 

To construct our theory first a quick review of integrating out a field when expanding on derivatives~\cite{Cohen:2020xca}. If one implements this expansion in the action for a UV complete model with heavy $\Phi$ and light $\phi$ fields
\begin{align}
S\left[\phi,\Phi\right]=&S^{(0)}\left[\phi,\Phi\right]+S^{(p^2)}\left[\phi,\Phi\right]+\dots \,,
\end{align}
and also expands the equation of motion for the heavy field $\Phi^{\textrm{EoM}}=\Phi^{\textrm{EoM}}_{(0)}+\Phi^{\textrm{EoM}}_{(p^2)}\dots$ whose first term satisfies 
\begin{align}
	 \frac{\delta S}{\delta \Phi}\left(\phi,\Phi^{\textrm{EoM}}(\phi)\right)&=\frac{\delta S^{(0)}}{\delta \Phi}\left(\phi,\Phi^{\textrm{EoM}}(\phi)\right)+\cdots=0 \,, & \frac{\delta S^{(0)}}{\delta \Phi}\left(\phi,\Phi^{\textrm{EoM}}_{(0)}(\phi)\right)&\equiv 0\label{EoM}\,,
\end{align}
one obtains an effective action at low energies and tree level
\begin{align}
S_{\textrm{eff}}(\phi)=&S[\phi,\Phi^{\textrm{EoM}}(\phi)]
\\=&S^{(0)}\left[\phi,\Phi^{\textrm{EoM}}_{(0)}(\phi)\right]+S^{(p^2)}\left[\phi,\Phi^{\textrm{EoM}}_{(0)}(\phi)\right]+\Phi^{\textrm{EoM}}_{(p^2)}\frac{\delta S^{(0)}}{\delta\Phi}\left[\phi,\Phi^{\textrm{EoM}}_{(0)}(\phi)\right]+\dots\\
=&S^{(0)}\left[\phi,\Phi^{\textrm{EoM}}_{(0)}(\phi)\right]+S^{(2)}\left[\phi,\Phi^{\textrm{EoM}}_{(0)}(\phi)\right]+\dots\,,
\end{align}
where we used the second equation in~(\ref{EoM}) in the last line.
It suffices then to solve the vanishing momenta equation of motion for the heavy field $\Phi$ and substitute in the order $p^2$ action to obtain the effective potential and kinetic terms at low energies and tree level.

We can now specify our model, which we take as simple as we can find it~\cite{Cohen:2020xca}. Take the Standard Model and add a singlet $S$, $(S)_G=S$ with a $Z_2$ symmetry and a coupling to a linear Higgs doublet $H$ as
 \begin{align}\label{Pot}
 	V(H,S)=-m_1^2H^\dagger H-\frac{m_2^2}{2}S^2-\frac{\tilde\lambda}{4} H^\dagger H S^2 +\frac{\lambda}{8} S^4+\frac{\lambda_H}{2} (H^\dagger H)^2\,,
 \end{align}
 with all parameters positive which implies in particular both $S$ and $H$ get a vev.
 
  To obtain our HEFT theory we integrate out $S$, which has
  a 0th order EoM
 \begin{align}
 	S\left(-m_2^2-\tilde\lambda H^\dagger H/2+\lambda S^2/2\right)=&0\,,& S^{\textrm{EoM}}_{(0)}&=\sqrt{\frac{2}{\lambda}\left( m_2^2+\tilde \lambda H^\dagger H/2 \right) }\label{SEoM}\,,
 \end{align}
 where we take the potential so as to produce a vev for $S$. It is good to keep an eye on the mass of the state we integrate out, we have with our approximations
 \begin{align}\label{SMass}
 	m_S^2=\left.\frac{\partial^2 V}{\partial S^2}\right|_{S=v_S}=\lambda v_S^2=2\left(\tilde \lambda \frac{v_\star^2}4+m_2^2\right)\,.
 \end{align}
 One obtains the effective action kinetic term subbing in~(\ref{SEoM})
 \begin{align}
 	\mathcal L_{\textrm{eff}}^{p^2}=&\mathcal L\left(S^{\textrm{EoM}}_{(0)}(H),H\right) =\frac12 \frac{2}{\lambda}\frac14 \frac{(\tilde\lambda\partial_\mu( H^\dagger H)/2)^2}{m_2^2+\tilde\lambda H^\dagger H/2}+D_\mu H^\dagger D^\mu H\\
 	=&\frac12\left[\frac{8m_2^2+\tilde\lambda(\tilde\lambda/\lambda+2)(v_\star+\underline h)^2}{8m_2^2+2\tilde\lambda( v_\star+\underline h)^2}\right](\partial\underline{h})^2+\frac{(v_\star+\underline h)^2}{4}\textrm{Tr}\left(D_\mu U^\dagger D^\mu U \right)\label{SQuot}\\
 	&\equiv \frac12 K(\underline h)(\partial \underline h)^2+\frac{v^2F(\underline h)^2}{4}\textrm{Tr}\left(D_\mu U^\dagger D^\mu U \right)\,,
 \end{align}
 as follows from using eq.~(\ref{HlinUh}). In view of the kinetic term of the Goldstones we realise that $v_\star=v$ which simplifies the following discussion.
 From here the curvature follows from a coordinate change and the chain rule, as you might have worked out yourself in the \textbf{DIY\#11} (with $L=vF=v+\underline h$) exercise, eq.~(\ref{RDefNew0}), and they read,
 \begin{align}\label{RDefNew}
 R_\varphi=&\frac{1}{v^2F^2}-\frac{(F')^2}{F^2}=\frac{1}{v^2F(\underline h)^2}\left(1-\frac{1}{K(\underline h)}\right),
 &
 	R_h=&-\frac{F''}{F}=\frac{K'(\underline h)}{2vF(\underline h)K^2(\underline h)}\,.
 \end{align}
 So one obtains after some work {\color{teal} ${||}$ which you can do yourself:} not much guidance needed, take eq.~(\ref{SQuot}) put in eq.~(\ref{RDefNew}) and compute {\color{teal} $||$}
 \begin{align}\label{RsS}
 	R_\varphi=&\frac{\tilde\lambda^2}{8\lambda m_2^2+\tilde\lambda(\tilde\lambda+2\lambda)(v+\underline h)^2},&
 	R_h&=\frac{8\tilde\lambda^2\lambda m_2^2}{\left(8\lambda m_2^2+\tilde\lambda(\tilde\lambda+2\lambda)(v+\underline h)^2\right)^2}\,.
 \end{align}
 Here we can see that this model interpolates between SMEFT and HEFT/SMEFT. Take the decoupling limit $m_2\gg v$ and you can check that the correlation of SMEFT to first order in $1/m^2_2$ is corroborated. However take $m_2=0$ --which does not yield a massless $S$ since there is the extra contribution in~(\ref{SMass})-- and we step outside of SMEFT and into HEFT/SMEFT;
the curvature $R_\varphi$ presents a pole at $\underline h=-v$ where it does blow up! The generalization of the LSZ formula tells us this curvature would be proportional to the amplitude when expanding around that point, so it is a true singularity. You can see what is the $R_h,R_\varphi$ correlation in this $m_2=0$ in eq.~(\ref{RsS}) limit to see if it differs from SMEFT. Finally one can push further into HEFT/SMEFT and take $m^2_2<0$ -- again still compatible with massive $S$ cf.~(\ref{SMass})-- this will also yield a pole, closer to $\underline h=0$ than $\underline h=-v$.

This simple model teaches us two important lessons.

\begin{itemize}
	\item  The first is that when we step into quotient theories, the mass of the particle we integrated out is tied down to the electroweak scale, in our case $m_S\leq (\tilde\lambda/2)^{1/2} v$, which means through the perturbative unitarity bound on $\tilde\lambda\leq 8\pi$, an upper bound on $m_S$. Indeed this theory was found not to decouple in~\cite{Cohen:2020xca} with a finite cut-off $\sim 4\pi v$ .
	\item The second is that in our trip towards HEFT/SMEFT, we first encounter the $m_2=0$ theory which has a candidate fixed point but that same point has divergent curvature; it is a type B theory. The $m^2_2<0$ theory on the other hand has the singularity before the would be fixed point but we can only trust our theory till the singularity. There is then no fixed point and we have a type A theory.
\end{itemize}

 The first point confirms the expectations of finite and $v$-related cut-off, i.e. non-decoupling theories where there is an upper bound on the mass of new states. This has been generalized for theories with a singularity making use of the Hadamard theorem to derive a finite cut-off of $\sim 4\pi v$~\cite{Cohen:2021ucp}. The argument goes something like this; if there is a singularity some distance $v_\star$ away from the vacuum, the Hadamard theorem tells us how fast the coefficients in the Taylor expansion on fields grow, these coefficients give us scattering and are in turn each bound by unitarity. Individual channels with n+2 particles point at a cut-off --derived as we did for Higgsless theory on eq.~(\ref{actU}) in the end of sec.~\ref{sec:Long}-- of $4\pi v_\star\sqrt{n}$, which we can expect to improve to $4\pi v_\star$ when summing over possible scattering in the inelastic channel. Finally as in the case above, the examples we know about have $v_\star\sim v$ and so we obtain a finite cut-off from unitarity. For a more detailed and rigorous exposition see~\cite{Cohen:2021ucp}.
 
\begin{figure}[h]
	\centering
	\includegraphics[width=.5\textwidth]{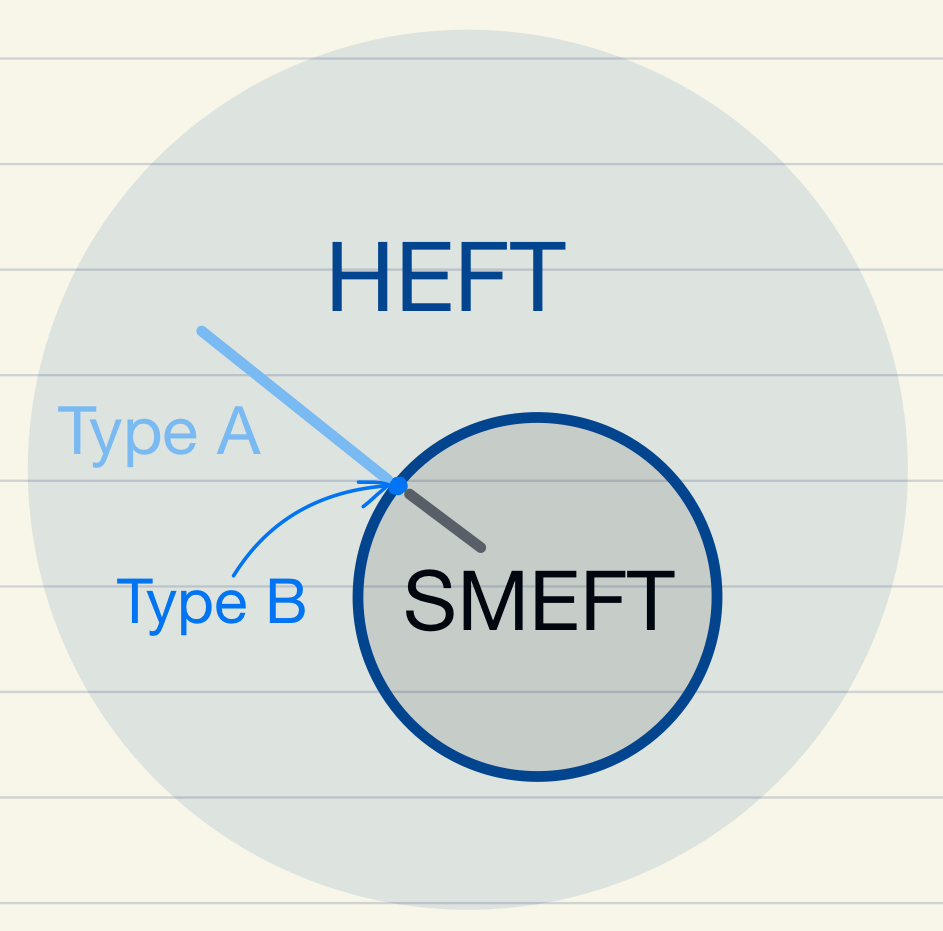}
	\caption{\label{Fig:probe}Postulated geography of quotient HEFT/SMEFT theories as deduced from our `trajectory' model in~(\ref{Pot}).}
\end{figure}

 The second point tells us about the geography of quotient HEFT/SMEFT theories, if we can extend our simple finding here, it seems type B theories lie in the boundary and beyond we find type A theories. This is sketched in figure~\ref{Fig:probe}.
 
   Let us end by noting that these results are good news for experiment, these type of theories are --$4\pi v\sim $TeV-- within reach!

\section{Phenomenology and legacy}\label{sec:PH}
 
 We have tried to show that HEFT is the most general EFT to describe the dynamics of the known particle spectrum; it allows for deviations from the SM in all theory directions provided any new physics is heavier than a few hundred GeV. We are exploring this electroweak theory space experimentally at LHC but there is still much we do not know about the Higgs sector, the triple Higgs coupling being an example of a property hard to probe.
 However we can make the best of what we know and compare with HEFT to obtain information on the coefficients, something that has been done in multiple articles, some are
 ~\cite{Brivio:2013pma,Brivio:2016fzo,deBlas:2018tjm}.

 The procedure to compare with data is the same as in SMEFT; we expand around the vacuum, find that the operators produce new vertexes, define our inputs and then make predictions for observables. The full list of new physics effects deriving from HEFT is extensive: electroweak physics at colliders, flavour physics for both quarks and leptons, baryon and number violation, CP violation and even low energy observables as a possible electric dipole moment of the neutron.
 
 For space constraints we cannot, nor aim to, cover even one of these sectors here. Instead let us note a couple of distinctive features of HEFT when compared to SMEFT. First, as noted earlier, HEFT allows for large deviations from the SM and in particular it can provide the theoretical background for the $\kappa$ framework of simply rescaling SM results. Second, a generic feature of HEFT is de-correlation of observables, understandable from the fact that it comprises SMEFT and exemplified by the curvature-correlation of first-order SMEFT of eq.~(\ref{Corr}) which is lost in HEFT.
 
 To put some numbers in our Lagrangian and get a sense of current precision, we borrow results from \cite{deBlas:2018tjm}, for a couple of coefficients
 \begin{align}
 	vF'(0)&=1.01 \pm 0.06 &
 	v\mathcal P'_{G_s}(0)\leq &\frac{g_s^2}{(4\pi)^2}(-0.01\pm 0.08) 
  \end{align}
  where $\mathcal P'_{G_s}$ controls the linear Higgs coupling to gluons and a prime denotes derivation wrt $h$. These parameters are accessed through linear couplings of the Higgs, vector boson or gluon fusion. Indeed while curvature is related to longitudinal scattering amplitudes, these are not easy to access experimentally and the strongest experimental constraints come from single Higgs production and --the absence of evidence for-- double Higgs production. This means in particular that the constraints on $R_h$ are weaker than on $R_\varphi$. These constraints on curvature are shown on fig.~\ref{Fig}.
  
 Finally let us come to the prominent feature of (we suspect  most if not all) quotient theories, a finite cut-off and their position potentially within reach of collider experiments.
 For this once more we turn to our example of eq.~(\ref{Pot}). I will let you do the work and use fig.~\ref{Fig} and what we have learned to do the phenomenological analysis outlined in the last DIY of these notes:
 
 \begin{figure}[h!]
 	\centering
 	\includegraphics[width=\linewidth]{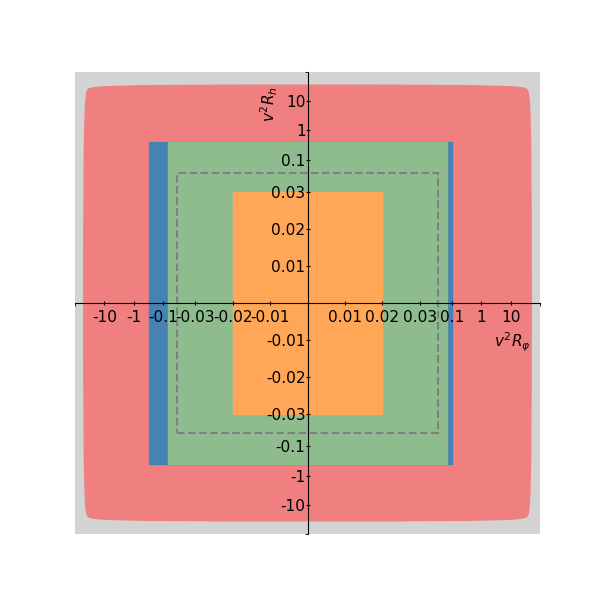}
 	\caption{\label{Fig} Experimental bounds on curvature; the red region is excluded by present LHC data, the LHC-HL would further explore/exclude the blue region and finally projected FCC sensitivity would narrow the range down to exploring the green region leaving only the organe inner rectangle in the unknown realm. Within the dashed square the scaling is logarithmic and outside is linear}
 \end{figure}

{\noindent\color{teal}{\rule{\textwidth}{1mm}}}
\textbf{Do it yourself \#12:} Testing your model. Inspect the two limits for the curvature

\begin{itemize}
\item  SMEFT, $m_2\gg v$. Identify the $R_h$ vs $R_\varphi$ correlation in this limit; is there any obstacle to taking  $R\to 0$?  
\item  HEFT/SMEFT $m_2^2\leq0$ Identify the $R_h$ vs $R_\varphi$ correlation in the limit $m_2=0$; is there any obstacle to taking $R\to 0$? To answer this you can compute the singlet $S$ mass subbing in $S=v_S+s$ and $H^\dagger H=v^2/2$ in the potential of eq.~(\ref{Pot}) to obtain eq.~(\ref{SMass}). How would the above results change for small negative $m_2^2$?
\end{itemize}
Put your findings on the plot of fig.~\ref{Fig} which presents circa 2022 and future experimental bounds. Is the quotient theory ruled out? Will it be by future experiments? You can take for numerics the requirement $m_S>2m_h$ for EFT to hold and from perturbative unitarity $\lambda,\tilde\lambda\leq 8\pi$.

 {\noindent\color{teal}{\rule{\textwidth}{1mm}}}

 \vspace{0.6cm}
 
\noindent\textbf{Closing remarks}\\

The LHC and future experiments will reveal the physics of the electroweak theory addressing questions like the dynamics behind gauge symmetry breaking, the origin of elementary particle masses, the hierarchy problem, and the dichotomy of the linear or non-linear nature of the scalar sector. No definite answer is guaranteed for the first three points, but there are encouraging theory results that make us think the last question \textit{will} be answered.

\section*{Acknowledgements}
I would like to thank the students of KIAS QUC 2022 school, Pyungwon Ko, Mia West, Madison Hammond and Dave Sutherland for feedback on these notes and the lectures.  I also acknowledge support from the STFC under Grant No. ST/T001011/1.

 \bibliographystyle{plain}
 \bibliography{HEFTLib}

\begin{thebibliography}{10}

\bibitem{Alonso:2012px}
R.~Alonso, M.~B. Gavela, L.~Merlo, S.~Rigolin, and J.~Yepes.
\newblock {The Effective Chiral Lagrangian for a Light Dynamical ''Higgs
  Particle''}.
\newblock {\em Phys. Lett. B}, 722:330--335, 2013.
\newblock [Erratum: Phys.Lett.B 726, 926 (2013)].

\bibitem{Alonso:2017tdy}
R.~Alonso, K.~Kanshin, and S.~Saa.
\newblock {Renormalization group evolution of Higgs effective field theory}.
\newblock {\em Phys. Rev. D}, 97(3):035010, 2018.

\bibitem{Alonso:2014wta}
Rodrigo Alonso, Ilaria Brivio, Belen Gavela, Luca Merlo, and Stefano Rigolin.
\newblock {Sigma Decomposition}.
\newblock {\em JHEP}, 12:034, 2014.

\bibitem{Alonso:2015fsp}
Rodrigo Alonso, Elizabeth~E. Jenkins, and Aneesh~V. Manohar.
\newblock {A Geometric Formulation of Higgs Effective Field Theory: Measuring
  the Curvature of Scalar Field Space}.
\newblock {\em Phys. Lett. B}, 754:335--342, 2016.

\bibitem{Alonso:2016oah}
Rodrigo Alonso, Elizabeth~E. Jenkins, and Aneesh~V. Manohar.
\newblock {Geometry of the Scalar Sector}.
\newblock {\em JHEP}, 08:101, 2016.

\bibitem{Alonso:2016btr}
Rodrigo Alonso, Elizabeth~E. Jenkins, and Aneesh~V. Manohar.
\newblock {Sigma Models with Negative Curvature}.
\newblock {\em Phys. Lett. B}, 756:358--364, 2016.

\bibitem{Alonso:2022ffe}
Rodrigo Alonso and Mia West.
\newblock {On the effective action for scalars in a general manifold to any
  loop order}.
\newblock 7 2022.

\bibitem{Alonso:2021rac}
Rodrigo Alonso and Mia West.
\newblock {Roads to the Standard Model}.
\newblock {\em Phys. Rev. D}, 105(9):096028, 2022.

\bibitem{Appelquist:1980vg}
Thomas Appelquist and Claude~W. Bernard.
\newblock {Strongly Interacting Higgs Bosons}.
\newblock {\em Phys. Rev. D}, 22:200, 1980.

\bibitem{Appelquist:1974tg}
Thomas Appelquist and J.~Carazzone.
\newblock {Infrared Singularities and Massive Fields}.
\newblock {\em Phys. Rev. D}, 11:2856, 1975.

\bibitem{Brivio:2013pma}
I.~Brivio, T.~Corbett, O.~J.~P. \'Eboli, M.~B. Gavela, J.~Gonzalez-Fraile,
  M.~C. Gonzalez-Garcia, L.~Merlo, and S.~Rigolin.
\newblock {Disentangling a dynamical Higgs}.
\newblock {\em JHEP}, 03:024, 2014.

\bibitem{Brivio:2016fzo}
I.~Brivio, J.~Gonzalez-Fraile, M.~C. Gonzalez-Garcia, and L.~Merlo.
\newblock {The complete HEFT Lagrangian after the LHC Run I}.
\newblock {\em Eur. Phys. J. C}, 76(7):416, 2016.

\bibitem{Brivio:2017vri}
Ilaria Brivio and Michael Trott.
\newblock {The Standard Model as an Effective Field Theory}.
\newblock {\em Phys. Rept.}, 793:1--98, 2019.

\bibitem{Buchalla:2020kdh}
G.~Buchalla, O.~Cat\`a, A.~Celis, M.~Knecht, and C.~Krause.
\newblock {Higgs-electroweak chiral Lagrangian: One-loop renormalization group
  equations}.
\newblock {\em Phys. Rev. D}, 104(7):076005, 2021.

\bibitem{Buchalla:2013rka}
Gerhard Buchalla, Oscar Cat\`a, and Claudius Krause.
\newblock {Complete Electroweak Chiral Lagrangian with a Light Higgs at NLO}.
\newblock {\em Nucl. Phys. B}, 880:552--573, 2014.
\newblock [Erratum: Nucl.Phys.B 913, 475--478 (2016)].

\bibitem{Buchalla:2013eza}
Gerhard Buchalla, Oscar Cat\'a, and Claudius Krause.
\newblock {On the Power Counting in Effective Field Theories}.
\newblock {\em Phys. Lett. B}, 731:80--86, 2014.

\bibitem{Callan:1969sn}
Curtis~G. Callan, Jr., Sidney~R. Coleman, J.~Wess, and Bruno Zumino.
\newblock {Structure of phenomenological Lagrangians. 2.}
\newblock {\em Phys. Rev.}, 177:2247--2250, 1969.

\bibitem{Cheung:2021yog}
Clifford Cheung, Andreas Helset, and Julio Parra-Martinez.
\newblock {Geometric soft theorems}.
\newblock {\em JHEP}, 04:011, 2022.

\bibitem{Cohen:2020xca}
Timothy Cohen, Nathaniel Craig, Xiaochuan Lu, and Dave Sutherland.
\newblock {Is SMEFT Enough?}
\newblock {\em JHEP}, 03:237, 2021.

\bibitem{Cohen:2021ucp}
Timothy Cohen, Nathaniel Craig, Xiaochuan Lu, and Dave Sutherland.
\newblock {Unitarity violation and the geometry of Higgs EFTs}.
\newblock {\em JHEP}, 12:003, 2021.

\bibitem{Cohen:2022uuw}
Timothy Cohen, Nathaniel Craig, Xiaochuan Lu, and Dave Sutherland.
\newblock {On-Shell Covariance of Quantum Field Theory Amplitudes}.
\newblock {\em Phys. Rev. Lett.}, 130(4):041603, 2023.

\bibitem{Coleman:1969sm}
Sidney~R. Coleman, J.~Wess, and Bruno Zumino.
\newblock {Structure of phenomenological Lagrangians. 1.}
\newblock {\em Phys. Rev.}, 177:2239--2247, 1969.

\bibitem{Contino:2013kra}
Roberto Contino, Margherita Ghezzi, Christophe Grojean, Margarete Muhlleitner,
  and Michael Spira.
\newblock {Effective Lagrangian for a light Higgs-like scalar}.
\newblock {\em JHEP}, 07:035, 2013.

\bibitem{Contino:2015mha}
Roberto Contino and Matteo Salvarezza.
\newblock {One-loop effects from spin-1 resonances in Composite Higgs models}.
\newblock {\em JHEP}, 07:065, 2015.

\bibitem{Cornwall:1974km}
John~M. Cornwall, David~N. Levin, and George Tiktopoulos.
\newblock {Derivation of Gauge Invariance from High-Energy Unitarity Bounds on
  the s Matrix}.
\newblock {\em Phys. Rev. D}, 10:1145, 1974.
\newblock [Erratum: Phys.Rev.D 11, 972 (1975)].

\bibitem{deBlas:2018tjm}
Jorge de~Blas, Otto Eberhardt, and Claudius Krause.
\newblock {Current and Future Constraints on Higgs Couplings in the Nonlinear
  Effective Theory}.
\newblock {\em JHEP}, 07:048, 2018.

\bibitem{Falkowski:2019tft}
Adam Falkowski and Riccardo Rattazzi.
\newblock {Which EFT}.
\newblock {\em JHEP}, 10:255, 2019.

\bibitem{Falkowski:2012vh}
Adam Falkowski, Slava Rychkov, and Alfredo Urbano.
\newblock {What if the Higgs couplings to W and Z bosons are larger than in the
  Standard Model?}
\newblock {\em JHEP}, 04:073, 2012.

\bibitem{Feruglio:1992wf}
F.~Feruglio.
\newblock {The Chiral approach to the electroweak interactions}.
\newblock {\em Int. J. Mod. Phys. A}, 8:4937--4972, 1993.

\bibitem{Gavela:2016bzc}
B.~M. Gavela, E.~E. Jenkins, A.~V. Manohar, and L.~Merlo.
\newblock {Analysis of General Power Counting Rules in Effective Field Theory}.
\newblock {\em Eur. Phys. J. C}, 76(9):485, 2016.

\bibitem{Gomez-Ambrosio:2022why}
Raquel G\'omez-Ambrosio, Felipe~J. Llanes-Estrada, Alexandre Salas-Bern\'ardez,
  and Juan~J. Sanz-Cillero.
\newblock {SMEFT is falsifiable through multi-Higgs measurements (even in the
  absence of new light particles)}.
\newblock 7 2022.

\bibitem{Graf:2022rco}
Luk\'a\v{s} Gr\'af, Brian Henning, Xiaochuan Lu, Tom Melia, and Hitoshi
  Murayama.
\newblock {Hilbert series, the Higgs mechanism, and HEFT}.
\newblock {\em JHEP}, 02:064, 2023.

\bibitem{Grinstein:2007iv}
Benjamin Grinstein and Michael Trott.
\newblock {A Higgs-Higgs bound state due to new physics at a TeV}.
\newblock {\em Phys. Rev. D}, 76:073002, 2007.

\bibitem{Helset:2022tlf}
Andreas Helset, Elizabeth~E. Jenkins, and Aneesh~V. Manohar.
\newblock {Geometry in scattering amplitudes}.
\newblock {\em Phys. Rev. D}, 106(11):116018, 2022.

\bibitem{Helset:2022pde}
Andreas Helset, Elizabeth~E. Jenkins, and Aneesh~V. Manohar.
\newblock {Renormalization of the Standard Model Effective Field Theory from
  geometry}.
\newblock {\em JHEP}, 02:063, 2023.

\bibitem{Lane:1993wz}
Kenneth~D. Lane.
\newblock {An Introduction to technicolor}.
\newblock In {\em {Theoretical Advanced Study Institute (TASI 93) in Elementary
  Particle Physics: The Building Blocks of Creation - From Microfermius to
  Megaparsecs}}, 6 1993.

\bibitem{Lee:1977eg}
Benjamin~W. Lee, C.~Quigg, and H.~B. Thacker.
\newblock {Weak Interactions at Very High-Energies: The Role of the Higgs Boson
  Mass}.
\newblock {\em Phys. Rev. D}, 16:1519, 1977.

\bibitem{Longhitano:1980iz}
Anthony~C. Longhitano.
\newblock {Heavy Higgs Bosons in the Weinberg-Salam Model}.
\newblock {\em Phys. Rev. D}, 22:1166, 1980.

\bibitem{Longhitano:1980tm}
Anthony~C. Longhitano.
\newblock {Low-Energy Impact of a Heavy Higgs Boson Sector}.
\newblock {\em Nucl. Phys. B}, 188:118--154, 1981.

\bibitem{Manohar:1983md}
Aneesh Manohar and Howard Georgi.
\newblock {Chiral Quarks and the Nonrelativistic Quark Model}.
\newblock {\em Nucl. Phys. B}, 234:189--212, 1984.

\bibitem{Sun:2022ssa}
Hao Sun, Ming-Lei Xiao, and Jiang-Hao Yu.
\newblock {Complete NLO Operators in the Higgs Effective Field Theory}.
\newblock 6 2022.

\bibitem{Vayonakis:1976vz}
C.~E. Vayonakis.
\newblock {Born Helicity Amplitudes and Cross-Sections in Nonabelian Gauge
  Theories}.
\newblock {\em Lett. Nuovo Cim.}, 17:383, 1976.

\bibitem{Workman:2022}
R.L. Workman et~al.
\newblock {Review of Particle Physics}.

\end{thebibliography}
\end{document}